\documentclass[hidelinks,a4paper,11pt]{article}

\usepackage[utf8]{inputenc}
\usepackage[english]{babel}
\usepackage[T1]{fontenc}

\usepackage[a4paper]{geometry}

\usepackage{mathtools}
\usepackage{amsthm, amssymb}
\usepackage{leftindex}
\usepackage{thmtools,thm-restate}
\usepackage[hidelinks]{hyperref}
\usepackage[nameinlink,noabbrev]{cleveref}
\usepackage{placeins}
\usepackage{graphicx}
\usepackage{adjustbox}
\usepackage{float}
\usepackage[skip=5pt, font=small]{caption}
\usepackage[shortlabels]{enumitem}
\usepackage{tikz}
\usepackage{pdflscape}
\usetikzlibrary{arrows.meta, positioning}

\usepackage{dsfont}
\usepackage{physics}
\usepackage{mleftright} %

\usepackage{booktabs}

\usepackage{natbib} 
\newcommand{\BR}{{\mathbb{R}}}

\newcommand{\CE}{{\cal E}}

\DeclareMathOperator*{\argmin}{argmin}

\DeclareMathSymbol{\shortminus}{\mathbin}{AMSa}{"39}

\DeclareMathOperator{\diag}{diag}

\newcommand{\Id}{\mathrm{Id}}

\DeclarePairedDelimiterX{\infdivx}[2]{(}{)}{%
  #1\;\delimsize\|\;#2%
}

\let\phi\varphi
\let\theta\vartheta
\let\epsilon\varepsilon

\let\le\leqslant
\let\ge\geqslant
\let\leq\leqslant

\newcommand\circlemarker{%
  \begin{picture}(1,1)
    \put(0.5,0.3){\circle*{0.55}} %
  \end{picture}%
}

\newcommand\squaremarker{%
  \begin{picture}(1,1)
    \put(0,0){\rule{0.55em}{0.55em}}
  \end{picture}%
}

\DeclareRobustCommand\xmarker{%
  \tikz[baseline=-0.3em, scale=0.15]{
    \draw[line width=0.25em] (-0.5,-0.5) -- (0.5,0.5);
    \draw[line width=0.25em] (-0.5,0.5)  -- (0.5,-0.5);
  }%
}

\author{Malte Londschien${}^{1,2,}$, Manuel Burger${}^{3}$, Gunnar R\"atsch${}^{3,4}$, and Peter B\"uhlmann${}^{1}$\\
\vspace{0.1cm}\\
{\small${}^{1}$Seminar for Statistics, ETH Z\"urich, Switzerland}\\
{\small${}^{2}$AI Center, ETH Z\"urich, Switzerland}\\
{\small${}^{3}$Department of Computer Science, ETH Z\"urich, Switzerland}\\
{\small${}^{4}$Swiss Institute for Bioinformatics, Z\"urich, Switzerland}
}
\title{Domain Generalization and Adaptation in Intensive Care with Anchor Regression}
\date{February 2026}
\begin{document}

\maketitle
\begin{abstract}
\noindent %
The performance of predictive models in clinical settings often degrades when deployed in new hospitals due to distribution shifts.
This paper presents a large-scale study of causality-inspired domain generalization on heterogeneous multi-center intensive care unit (ICU) data. We apply anchor regression and introduce anchor boosting, a novel, tree-based nonlinear extension, to a large dataset comprising 400,000 patients from nine distinct ICU databases.
We find that anchor regularization yields improvements of
out-of-distribution performance, particularly for the most dissimilar target domains.
The methods appear robust to violations of theoretical assumptions, such as anchor exogeneity.
Furthermore, we propose a novel conceptual framework to quantify the utility of large external data datasets.
By evaluating performance as a function of available target-domain data, we identify three regimes: (i) a domain generalization regime, where only the external model should be used, (ii) a domain adaptation regime, where refitting the external model is optimal, and (iii) a data-rich regime, where external data provides no additional value.

\end{abstract}

\normalsize

\vspace{0.2cm}
\noindent\small {Accepted for publication in RSS: Data Science and Artificial Intelligence.}
\normalsize
\section{Introduction}

A standard assumption in predictive modeling is that training and test data come from the same distribution.
This assumption often fails in real-world scenarios.
For example, in clinical applications, test data may originate from a different time period or hospital than the training data.
When these distribution shifts occur, model performance tends to drop significantly \citep{barak2021prediction,guo2021systematic,roland2022domain,yang2022machine,water2023yet,huser2024comprehensive}.

The field of distributional robustness has emerged to address this problem, but its successes have been largely demonstrated on simulated, semi-synthetic (for example, colored MNIST of \citealp{arjovski2019invariant} and ImageNet-C of \citealp{hendrycks2019benchmarking}), or curated (for example, PACS of \citealp{li2017deeper} and waterbirds of \citealp{sagawa2019distributionally}) datasets.
In contrast, large-scale empirical studies show mixed or negative results, with domain generalization models often failing to outperform simple baselines \citep{gulrajani2020search,guo2022evaluation,rockenschaub2024impact}.
An alternative approach to achieve generalization is to scale data and model capacity, a strategy proven successful for large language models \citep{brown2020language}. This is the focus of prior work by \citet{burger2025foundation}, who develop a foundation model on large ICU data.
They establish a ``square-root'' scaling law for domain generalization, where quadrupling the external data provides a similar performance gain to doubling the locally available data.

In contrast, we focus here on methods that use existing heterogeneity %
to improve robustness by exploiting causal models.
In particular, we consider anchor regression \citep{rothenhausler2021anchor}.
Intuitively, we expect causal relationships (vasopressor drugs raise blood pressure) to be stable, whereas relationships induced through hidden confounding (clinicians prescribe vasopressor drugs to severely ill patients, and thus vasopressor drug use correlates with increased mortality) can shift with varying treatment policies.
Anchor regression promotes stability or invariance to such shifts by penalizing dependencies that vary with the so-called anchor variable.
It achieves this by interpolating between ordinary least-squares and instrumental variables regression.

In this paper, we apply and extend anchor regression for medical prediction.
Recognizing that linear models may be insufficient to capture the complex feature interactions in clinical data, we first propose a novel, nonlinear extension to anchor regression based on gradient boosting trees \citep{friedman2001greedy}, popular in clinical predictive modeling \citep{hyland2020early,lyu2024empirical,huser2024comprehensive}.
We then conduct a large-scale empirical study, applying linear anchor regression and our nonlinear extension to predict adverse events in intensive care units (ICUs).
We assess the method's ability to improve distributional robustness on a strongly heterogeneous dataset aggregating nine ICU databases \citep{burger2024foundationmodelscriticalcare,burger2025foundation}.
Finally, we propose a conceptual framework for quantifying the utility of large external datasets, particularly when target domain data is scarce.

\subsection{Our contribution}
Our contributions are threefold:

\paragraph{A novel nonlinear extension to anchor regression}
We introduce \emph{anchor boosting}, a novel nonlinear extension of anchor regression \citep{rothenhausler2021anchor}  based on boosted tree learners.
While the concept of anchor boosting had been proposed before \citep{buhlmann2020invariance}, our implementation extends to classification tasks \citep{kook2022distributional} and incorporates second-order optimization to update tree leaf node values.
These extensions prove to be crucial for our application.

\paragraph{A large-scale application of causality-inspired regularization}
We conduct the largest-scale application of anchor regression \citep{rothenhausler2021anchor} and its variants to date, using a dataset of  400,000 patients and 10 million observations from nine distinct ICU databases.
To our knowledge, this work is also the first application of anchor regression and the largest application of a causality-inspired method to a medical prediction problem.
In a setting where many other domain generalization methods have been shown to provide little benefit over simple baselines \citep{rockenschaub2024impact}, we show that the anchor methods yield notable performance improvements, particularly for the most out-of-distribution target domains.

\paragraph{A framework to quantify the value of external data}
We propose and empirically validate a framework, with thematic similarities to \citet{desautels2017using}, to assess the utility of external data for a given target domain and task.
By expressing performance as a function of target sample size, we identify three regimes: (i) a domain generalization regime where only external data should be used, (ii) a domain adaptation regime, where refitting an external model is optimal, and (iii) a data-rich regime, where training on target data is best and the external data provides no additional value.
This taxonomy provides a practical methodology to quantify the information value of large external datasets in terms of equivalent number of in-domain samples.

\section{Data description}
\label{sec:data}
We use data from \citet{burger2024foundationmodelscriticalcare,burger2025foundation} who build upon the R-package \texttt{ricu} \citep{bennett2023ricu} to harmonize and aggregate ICU data from different sources.
We will describe this dataset in further detail next.

\subsection{ICU datasets included}
\label{sec:which_datasets}
We focus on the following 9 ICU datasets:
The \textbf{eICU Collaborative Research Database} \citep{pollard2018eicu} is a multi-center critical care database of $207$ hospitals throughout the USA. It contains data from 188'257 patient stays from 2015 and 2016.
The \textbf{Medical Information Mart for Intensive Care (MIMIC) III} \citep{johnson2016mimiciiia}
and \textbf{IV} \citep{johnson2023mimiciv}
contain data from the Beth Israel Deaconess Medical Center (BIDMC) in Boston (USA).
The BIDMC switched critical care information systems in 2008 from Philips CareVue to iMDsoft Metavision.
We consider only the CareVue subset of MIMIC-III to avoid an overlap with MIMIC-IV.
This MIMIC-III CareVue subset contains data from 34'154 patient stays from 2001--2008 and includes neonatal patient stays.
MIMIC-IV contains data from 93'679 patient stays from 2008--2022.
The \textbf{Northwestern ICU database (NWICU)} \citep{moukheibert2024northwestern} is a multi-center critical care database of 12 hospitals around Chicago (USA).
It contains data from 28'150 patient stays from 2020--2022.
The \textbf{High-Resolution ICU dataset (HiRID)} \citep{hyland2020early},
\textbf{Amsterdam University Medical Center database (AUMCdb)} \citep{thoral2021sharing}, and %
\textbf{Salzburg Intensive Care database (SICdb)} \citep{rodemund2024sicdb},
are single-center European datasets.
HiRID contains data from 33'586 patient stays from 2008--2016.
AUMCdb contains data from 22'897 patient stays from 2003--2016.
SICdb contains data from 27'115 patient stays from 2013--2021.
The \textbf{Paediatric Intensive Care database (PICdb)} \citep{li2019paediatric} contains data from the Children's Hospital of Zhejiang University School of Medicine in China.
It contains data from 13'516 patient stays from 2010--2018.
The \textbf{Critical Care Database Comprising Patients With Infection at Zigong Fourth People's Hospital} \citep{xu2022critical}
in China contains data of patients with a suspected infection.
It contains data from 2'583 patient stays from 2019 and 2020.
All datasets except for AUMCdb are available on PhysioNet \citep{goldberger2000physiobank}.
We summarize these data in \cref{tab:dataset_summaries} in appendix \ref{app:data}.

\subsection{Which type of information is measured in the ICU}

The variables measured in the ICU can be roughly divided into five categories: (i) patient demographics, (ii) vital signs, (iii) laboratory test results, (iv) treatments, and (v) auxiliary information.
(i) Patient demographics are variables that are assumed to be constant over a patient's stay, including biological sex, age, weight, and height.
(ii) Vital signs are continuously monitored variables used to assess a patients vital functions, including heart rate, blood pressure, and body temperature.
(iii) Laboratory tests measure the abundance of various substances in blood, including metabolism indicators such as creatinine and lactate.
(iv) Treatments are actions taken by the medical staff to treat a patient, including the administration of drugs, oxygen through ventilation, or other substances such as electrolytes and fluids.
(v) Finally, auxiliary information include the type of admission, the (approximate) year of admission, and a hospital ward identifier.
We use these variables to define environments or so-called anchors encoding heterogeneity in the data, as explained in \cref{sec:results:which_anchor}.

\subsection{Patient outcomes}
\label{sec:data:outcomes}
Research on the prediction of adverse events in the ICU is extensive \citep{tomavsev2019clinically,hyland2020early,yeche2021hirid,moor2023predicting,lyu2024empirical,rockenschaub2024impact,huser2024comprehensive}.
Although there is a conceptual consensus on how to define clinical events,
studies vary in their treatment of missing values and the logic used to convert clinical diagnostic event annotations into early event prediction labels.
This makes it difficult to compare performance scores across studies.
We follow simpler methods from the literature that are expected to generalize better across datasets.

We focus on four tasks: binary early event prediction (EEP) for circulatory failure and acute kidney failure and the corresponding continuous regression tasks of predicting log(lactate) and log(creatinine) levels.

We first define the underlying clinical events.
A patient is experiencing circulatory failure if they have low blood pressure (mean arterial pressure below 65 mmHg or receiving treatment to elevate it) and high blood sample lactate (above 2 mmol/l).
Acute kidney injury (AKI) is defined as AKI stage 3 according to the KDIGO guidelines \citep{kidney2012kdigo}.
Roughly, this means that a patient has high creatinine levels, low urine output, or is receiving renal replacement therapy.

From these diagnostic event annotations, we derive the prognostic binary early event prediction labels as follows:
(i)~If there is a positive event at the current time step, the label is missing.
(ii)~If the last event in the patient's history was positive, and there is a positive event within the forecast horizon, the label is missing.
(iii)~If there was no positive event in the patient's history or the last event in the patient's history was negative, and there is a positive event within the forecast horizon, the label is true.
(iv)~Else, if there is a negative event within the forecast horizon, the label is false.
This logic ensures that only switches between stable and unstable are considered.

For the binary prediction tasks, we follow prior work of \citet{hyland2020early} and \citet{lyu2024empirical} and use forecast horizons of 8 hours for circulatory failure and 48 hours for acute kidney injury.
These horizons reflect the timescales at which these organ systems degrade.
For the corresponding regression tasks, we use half of these horizons.

We visualize the resulting outcome distributions in \cref{fig:outcomes} and provide summary statistics in \cref{tab:dataset_summaries} in appendix \ref{app:data}.
See also \href{https://github.com/eth-mds/icu-features}{\texttt{github.com/eth-mds/icu-features}}.
\begin{figure}[htbp]
    \centering
    \includegraphics[width=\textwidth]{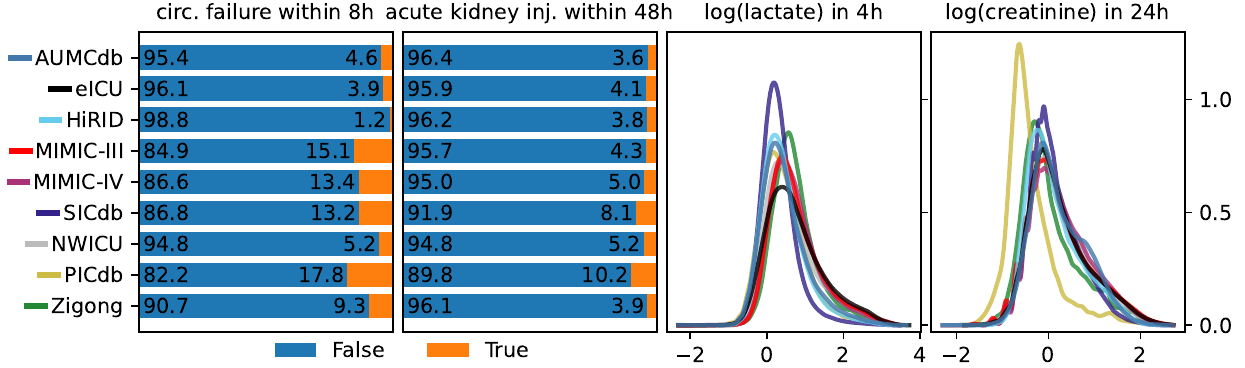}
    \caption{
    Distributions of binary and continuous outcomes. 
    }
    \label{fig:outcomes}
\end{figure}

\subsection{Feature engineering and pre-selection of variables}
The hourly time-series data, extracted by \citet{burger2024foundationmodelscriticalcare,burger2025foundation} using \texttt{ricu} \citep{bennett2023ricu} must be transformed into a static feature set to be used with standard linear or tree-based methods.
Following the literature, we engineer features over a backwards looking time window 
to capture some temporal dynamics of a patient's history.
We use an 8 hour horizon for circulatory failure and log(lactate) prediction and a 24 hour horizon for acute kidney injury and log(creatinine) prediction.
These features include a missingness indicator, the mean, standard deviation, maximum, minimum, and a linear trend for continuous variables, the mode and a missingness indicator for categorical variables, and an indicator and the average rate for treatments.
See \href{https://github.com/eth-mds/icu-features}{\texttt{github.com/eth-mds/icu-features}} and appendix \ref{app:data:feature_engineering} for details.

For computational reasons and to reduce model complexity, we do not use all available variables of \citet{burger2024foundationmodelscriticalcare,burger2025foundation}'s export in our models.
For the prediction of circulatory failure and log(lactate), we use the top 20 variables of \cite{hyland2020early} according to their table 1.
For the prediction of acute kidney injury and log(creatinine), we use the top variables of \citet{lyu2024empirical} according to their figure 8a.
See appendix \ref{app:data:variable_selection} for details.
Together with feature engineering, this results in 100--200 covariates for each task.

\subsection{Sources of heterogeneity}
\label{sec:data:heterogeneity}
We would like to emphasize the strong heterogeneity of the dataset at hand.
The dataset sources span three continents.
Possible sources of distribution shifts include:
(i) Different hardware and software used to measure and store vital signs and lab values. For example, Philips CareVue and iMDSoft Metavision for MIMIC-III (CV) and MIMIC-IV, respectively.
(ii) Different hospital policies. For example, higher willingness to prescribe certain medication or higher frequencies of lab value measurements.
(iii) Different cohort selection. For example, all databases except MIMIC-III and PICdb exclude non-adults, and the Zigong EHR database only includes patients with a suspected infection.
(iv) Different availability of variables.
For example, PICdb and Zigong only include very sparse measurements of mean arterial pressure, relevant for the diagnosis of circulatory failure, and NWICU has a lower data density compared to the other datasets.

Because of the cohort selection and variable availability, we expect the largest differences in cohort composition and variable availability for NWICU, PICdb, and Zigong.
We therefore designate these as ``truly out-of-distribution'' datasets.
In \cref{sec:results}, except for \cref{sec:results:refitting,sec:results:regimes}, we will use these only for evaluation and not for training. We believe that this reflects a realistic use-case, where a practitioner targeting a general adult ICU would likely exclude data from very dissimilar cohorts, such as the pediatric population in PICdb, from the training set.

\section{Methods}
This section details our methodological approach.
We begin by introducing linear anchor regression \citep{rothenhausler2021anchor}.
We then propose anchor boosting, a novel, nonlinear extension based on gradient boosting trees.
Finally, we describe our refitting procedure and introduce a taxonomy to quantify the value of external data.

\subsection{Linear anchor regression}
\label{sec:methods:anchor_regression}
The anchor regression estimator, proposed by \citet{rothenhausler2021anchor}, provides distributional robustness guarantees by guarding against potentially strong perturbations in a so-called anchor regression structural causal model.
The method is related to instrumental variables regression, where exogenous instruments $A \in \BR^{n \times k}$ that only affect the treatment variables of interest encode heterogeneity in the data, enabling causal effect estimation.
Anchor regression relaxes the strong assumption that the instruments, or anchors, do not directly affect the outcome or hidden confounders.
This comes at the cost of losing the possibility for inferring causality, but the method's causality inspired invariance regularization achieves distributional robustness.
It is exactly this robustness property which is useful in domain adaptation.

Consider a discrete anchor variable given by environments $e \in \CE$, for example, the ICU datasets $\CE = \{ \mathrm{AUMCdb},\  \mathrm{eICU}, \ \ldots, \ \mathrm{SICdb}\}$.
The linear anchor regression estimator is defined as
\begin{equation}
    \label{eq:anchor}
    \hat\beta_\mathrm{anchor}(\gamma) = \argmin_\beta \sum_{e\in\CE} \Bigg[ \sum_{i \in e} (y_i - X_i \beta)^2 + (\gamma - 1) \cdot \frac{1}{n_e} \Big(\sum_{i\in e} y_i - X_i \beta\Big)^2 \Bigg],
\end{equation}
where $\gamma \ge 1$ is the invariance regularization parameter and $n_e$ are the number of samples from environment $e$.
That is, anchor regression penalizes differences in the environments' mean residuals.
From a theoretical perspective, the anchor regression estimator optimizes a worst-case risk over a set of new, unseen environments $e \notin \CE$, for example $e = \mathrm{PICdb}$, with distribution shifts similar to the heterogeneity seen within the training data, but of a larger magnitude scaled by $\gamma$.

More generally, for possibly continuous anchor variables, write $P_A := A (A^T A)^{-1} A^T$ for the linear projection matrix onto the column space of $A$, such that $P_A \cdot v$ are the predictions of the linear model regressing $v \in \BR^n$ on $A \in \BR^{n \times k}$.
Then, the anchor regression estimator is
\begin{equation}
    \label{eq:anchor_general}
    \hat\beta_\mathrm{anchor}(\gamma) =
    \argmin_\beta \| y - X \beta \|_2^2 + (\gamma - 1) \cdot \| P_A (y - X \beta) \|_2^2.
\end{equation}
In practice, particularly in higher dimensions, we add an elastic-net regularization term $\lambda \Big( \eta  \|\beta\|_1 + (1 - \eta ) \|\beta\|_2^2 \Big)$ to \cref{eq:anchor,eq:anchor_general}, with $\lambda > 0$ and $0 \le \eta \le 1$.
We investigate the importance of such regularization in \cref{sec:results:hyperparameters}.

\subsection{Nonlinear anchor boosting (regression)}
\label{sec:anchor_boosting_regression}
The theoretical robustness guarantees of anchor regression mentioned above apply for a linear model \citep{rothenhausler2021anchor} or when using a nonlinear embedding \citep{sola2025causality}.
Such guarantees are unavailable for the proposed anchor boosting, except for some very preliminary result on risk constancy with full invariance corresponding to $\gamma = \infty$ \cite[proposition 5.1]{buhlmann2020invariance}.
Nonlinear methods generally cannot reliably extrapolate to shifted regimes \citep{Christiansen2022Causal}.
While linear models may diverge in such settings (correctly, if the model is true), tree ensembles inherently predict constant values outside the training range. This keeps extrapolation of prediction in the range of observed values, thereby handling distribution shifts conservatively.
Relying on this stability and the intuition from the linear case that homogenizing residuals across environments promotes robustness, we optimize \cref{eq:anchor,eq:anchor_general} by replacing $X_i \beta$ with a nonlinear function $f(X_i)$.

We focus here on boosted tree learners \citep{friedman2001greedy}, a popular method for ICU data \citep{hyland2020early,lyu2024empirical}. 
Analogously to \eqref{eq:anchor_general}, let
\begin{equation}
    \label{eq:nonlinear_anchor_loss}
    \ell(f, y) := \frac{1}{2} \| y - f \|^2 + \frac{1}{2} (\gamma - 1) \cdot \| P_A (y - f) \|^2
\end{equation}
be the anchor loss with gradient and Hessian with respect to $f$ as follows:
\begin{equation*}
    g(f, y)  = - (y - f) - (\gamma - 1) P_A (y - f) \quad \text{and} \quad H(f,y) = \Id + (\gamma - 1) P_A.
\end{equation*}
Following the construction of gradient boosting \citep{friedman2001greedy}, let $\hat f^j$ be the boosted learner after $j$ steps of boosting.
We initialize with $\hat f^0 := \frac{1}{n} \sum_{i=1}^n y_i$.
We then fit  the negative gradient against $X$ using a decision tree
$\hat t^{j+1} := - g(\hat f^j(X), y) \sim X$.
Let $M \in \BR^{n \times \mathrm{num. \ leafs}}$ be the one-hot encoding of $\hat t^{j+1}(X)$'s leaf node indices.
Then, $M^T \, g(\hat f^j(X), y)$ and $M^T \,H(\hat f^j(X), y) \, M$ are the gradient and Hessian of the loss function $\ell(\hat f^j(X) + \hat t^{j+1}(X), y) = \ell(\hat f^j(X) + M \hat\beta^{j+1}, y)$ with respect to $\hat t^{j+1}$'s leaf node values $\hat\beta^{j+1} \in \BR^{\mathrm{num. \ leafs}}$.
We set them using a second order optimization step to
$$
\hat \beta^{j+1} = - \left( M^T \, H(\hat f^j(X), y) \, M \right)^{-1} M^T \, g(\hat f^j(X), y).
$$
As the anchor regression loss \eqref{eq:nonlinear_anchor_loss} is quadratic in $f$, this second order optimization step actually yields the global optimum $\hat \beta^{j+1} = \argmin_\beta \ell(\hat f^j(X) + M \beta, y)$.
Finally, we set $\hat f^{j+1}(\cdot) := \hat f^j(\cdot) + \mathrm{lr} \cdot \hat t^{j+1}(\cdot)$, where the learning rate $\mathrm{lr}$ is typically set to $\mathrm{lr} = 0.1$. 

Our implementation of anchor boosting builds upon LightGBM \citep{ke2017lightgbm}.
It can be found at \href{https://github.com/mlondschien/anchorboosting}{\texttt{github.com/mlondschien/anchorboosting}}.
The implementation is very fast, taking between 30 and 60 seconds to train a 1'000 tree boosted anchor regression model on 1'000'000 observations with 100 features using 6 environments as anchor on a machine with 32 CPU cores.

A nonlinear extension of anchor regression based on boosting was already proposed by \citet{buhlmann2020invariance}.
However, this proposal does not use second order optimization to update the leaf node values, which appears to be crucial when using larger values of $\gamma$, especially for classification as described in \cref{sec:anchor_boosting_classification}.
See appendix \ref{app:second_order_tree_node_value_optimization} for details.
In related work, \citet{ulmer2025spectrally} fit a random forest to a linear rescaling of the data.
In \citet{ulmer2025anchorforest} they discuss how this can be applied to anchor regression, resulting in split points for each tree chosen to directly minimize the anchor loss.
However, this approach increases computational complexity beyond a near-linear relationship with sample size, making it prohibitive for the large datasets that we address in this paper.

\subsection{Nonlinear anchor boosting (classification)}
\label{sec:anchor_boosting_classification}
We apply
anchor regularization to binary classification tasks.
\citet{kook2022distributional} suggest to use the gradient of the log-likelihood as score residuals.
We use a probit link function, as, in contrast to the logistic link, the resulting anchor classification objective is convex. This convexity is necessary for stable second-order updates of the tree-leaf values.
See appendix \ref{app:logistic_anchor_is_non_convex} for details.

Write $\Phi$ and $\phi$ for the Gaussian distribution's cumulative distribution function and probability density function.
For binary classification with scores $f \in \BR^n$ and $y \in \{-1, 1\}^n$, the negative log-likelihood is $- \sum_{i=1}^n \log(\Phi(y_i f_i))$ with negative
gradient $ r := y \cdot \phi(f) / \Phi(y f)$.
We thus apply the same procedure as in \cref{sec:anchor_boosting_regression} to the loss
\begin{equation}
    \label{eq:probit_loss}
    \ell(f, y) := - \sum_{i=1}^n \log(\Phi(y_i f_i)) + \frac{1}{2} (\gamma - 1) \cdot \| P_A r \|^2,
\end{equation}
replacing the squared error $\frac{1}2(y_i - f_i)^2$ in \cref{eq:nonlinear_anchor_loss} with the negative log-likelihood $-\log(\Phi(y_i f_i))$ and the residuals $y_i - f_i$ with the likelihood score residuals $r_i$. 
We initialize with 
$\hat f^0 = \Phi^{-1}(\frac{1}{n} \sum_{i=1}^n \frac{y_i + 1}{2})$.
See appendix \ref{app:probit_anchor_loss} for derivations of the gradient and Hessian of \cref{eq:probit_loss}.

\subsection{Refitting using few target samples (linear models)}
\label{sec:methods:refitting}
In addition to the standard out-of-distribution (or domain) generalization setting, we consider the domain adaptation setting where some samples $(X_i, y_i)_{i \in e_\mathrm{target}}$ from the target environment are available.

Our goal is to effectively combine the external (source) data with the limited data from the target environment.
We use the external data to estimate a prior distribution, enabling a Bayesian approach to prediction in the target domain.
One natural approach is via empirical Bayes, assuming Gaussian target data and a Gaussian prior centered around $\hat\beta_\mathrm{source}$, the parameter estimate obtained from anchor regression trained on the source data.
The resulting maximum a posteriori estimate is then given by
\begin{equation}
\label{eq:emp_bayes}
\hat\beta_\mathrm{emp.\, Bayes} := \argmin_\beta \sum_{i \in e_\mathrm{target}} (y_i - X_i \beta )^2 + \alpha \| \beta - \hat\beta_\mathrm{source} \|^2_2,
\end{equation}
where the hyperparameter $\alpha$ controls the trade-off between the prior's and the target data's influence.
We also add an elastic net penalty regularizer, as written below \cref{eq:anchor_general}.
We jointly select $\alpha$ and $\hat\beta_\mathrm{source}$'s tuning parameters with 5-fold cross-validation on the target data.
For classification, we replace the squared error loss with the binomial negative log-likelihood.

\subsection{Refitting using few target samples (boosted tree models)}
\label{sec:methods:refitting_nonlinear}
We adapt the empirical Bayes estimator in \eqref{eq:emp_bayes} to nonlinear boosting tree algorithms.
For this, we use a pre-trained anchor boosting model from external data and update its tree's leaf node values using the new target data.

Set $\hat f^0_\mathrm{refit} = \hat f^0$.
Starting from $\hat f^j_\mathrm{refit} \in \BR^{| e_\mathrm{target} |}$, we drop the target data down the tree $\hat t^{j+1}$'s structure.
Using a loss with $\gamma=1$ (no invariance regularization), let $\hat\beta^{j+1}_\mathrm{new}$ be the second order optimization of the loss $\ell(\hat f_\mathrm{refit}^j + \hat t^{j+1}(X_\mathrm{target}), y_\mathrm{target})$ on the target data with respect to the leaf node values $\hat\beta^{j+1}$ of $\hat t^{j+1}$.
If there were no target samples in leaf node $k$, we set $(\hat \beta^{j+1}_{\mathrm{new}})_k = (\hat \beta^{j+1}_{\mathrm{old}})_k$.
Finally, we set $\hat\beta^{j+1}_\mathrm{refit} = \mathrm{dr} \cdot \hat\beta^{j+1}_\mathrm{old} + (1 - \mathrm{dr}) \cdot \hat\beta^{j+1}_\mathrm{new}$.
The decay rate $\mathrm{dr}$ is a tuning parameter, similar to $\alpha$ in \cref{eq:emp_bayes}.
We jointly select $\hat f_\mathrm{source}$'s only tuning parameter $\gamma$ and the decay rate $\mathrm{dr}$ via 5-fold cross-validation on the target data.

Thus, we refit the individual tree's leaf values but not the tree's structure, given by its split variables and thresholds.
This limit on the model's flexibility is advantageous when target data is scarce, as updating leaf node values requires fewer samples than learning split thresholds.
With abundant target data, we expect a model trained from scratch to eventually achieve superior performance.

This matches LightGBM's \citep{ke2017lightgbm} \texttt{refit} mechanism, except that (i) we use a probit link for classification, (ii) we use $\hat f^0_\mathrm{refit} = \hat f^0$ instead of re-estimation from the target data, and (iii) we do not update tree node values without any samples from the target instead of shrinking them towards zero.

\subsection{The value of external data and three regimes}
\label{sec:methods:regimes}
We consider prediction performance as a function of available samples or patients from the target domain, a perspective that shares themes with \citet{desautels2017using}.
We ask: How many target samples or patients are necessary to achieve a certain performance?
This can be used to quantify the value of large external data for a certain target domain, and, inversely, to quantify how far out-of-distribution the target domain lies.
This leads to a methodological taxonomy for describing heterogeneous datasets and domain adaptation.

\begin{figure}[tbph]
    \centering
    \hspace*{\fill}
    \includegraphics[width=0.95\textwidth]{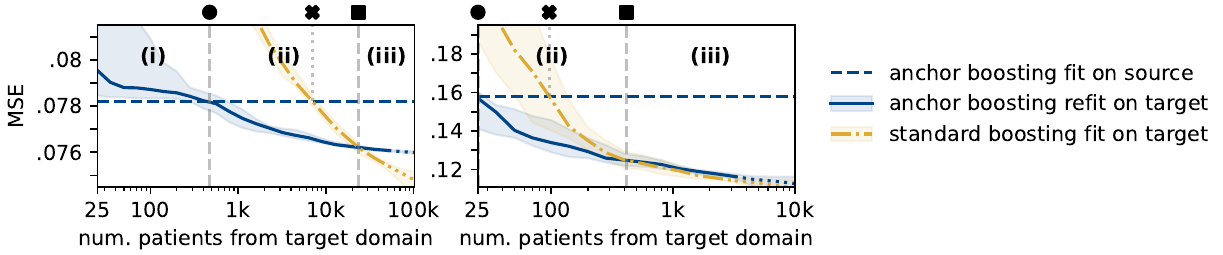}
    \caption{
        \label{fig:illustration}
        MSE predicting log(creatinine) in 24 hours as a function of available patients from the target domains eICU (left) and PICdb (right).
    }
\end{figure}

\Cref{fig:illustration} illustrates this taxonomy, where we compare the performance of an anchor boosting model as described in \cref{sec:anchor_boosting_regression,sec:anchor_boosting_classification}, anchor boosting models refit on target data as described in \cref{sec:methods:refitting_nonlinear}, and a regular boosted tree model fitted on the available target data only, predicting log(creatinine) in 24 hours on eICU and PICdb.
There are three regimes:
(i) If very few samples from the target distribution are available, it is best to use a model that was trained on source data only, including for model selection (domain generalization);
(ii) If more samples from the target distribution are available, it is best to use them to refit a model trained on the source data (domain adaptation);
(iii) If a large number of samples from the target distribution are available, it is best to ignore the source data and train a model on the target data only.

The line intersections in \cref{fig:illustration} carry the following interpretation:
\circlemarker \ The number of patients from the target domain to which using them for modeling does not improve performance.
\xmarker \ The value of external data for the target domain of interest.
\squaremarker \ The number of patients from the target domain from which one should ignore external data.

We present and discuss more figures similar to \cref{fig:illustration} in \cref{sec:results:refitting}.
We summarize all such plots for different tasks and target domains in \cref{fig:regimes} in \cref{sec:results:regimes}.

\section{Results}
\label{sec:results}
We apply nonlinear anchor boosting and linear anchor regression to the ICU data described in \cref{sec:data}.
Unless stated otherwise, the anchor boosting models use LightGBM's default values for hyperparameters, except for individual tree's maximal depth, which we restrict to 3, the total number of trees, which we increase to 1000, and the minimal gain to split, which we set to 0.1 to avoid 
splitting nodes with zero variance.
Limiting the maximum depth is recommended, as it drastically reduces the variance of LightGBM's leaf-wise tree growth algorithm.
We discuss the effect of hyperparameters on anchor boosting's out-of-distribution performance in \cref{sec:results:hyperparameters}.

We divided all data sets along patient identifiers into a 85\% train and a 15\% test set.
We designate AUMCdb, eICU, HiRID, MIMIC-III (CareVue subset), MIMIC-IV, and SICdb as core datasets, and NWICU, PICdb, and Zigong as ``truly out-of-distribution'' (OOD).
Performances shown for the core datasets result from models that were trained on the remaining 5 core datasets.
Performances shown for the truly OOD datasets result from models that were trained on all 6 core datasets.
The training sets were only used to train models and the test sets were only used to evaluate models and algorithms.

Except for \cref{sec:results:which_anchor}, where we study the effect of using different variables as anchor, we use the discrete dataset ID (AUMCdb, eICU, ...) as the anchor.
In \cref{sec:results:refitting,sec:results:regimes} we analyze the effect of refitting linear and boosted models on small samples from the target.

\subsection{Anchor regularization improves generalization to some ICU datasets}
\label{sec:results:1}
We observe that the causality-inspired regularization of nonlinear anchor boosting and linear anchor regression improves generalization to new ICU datasets.

\begin{figure}[ht]
    \includegraphics[width=\textwidth]{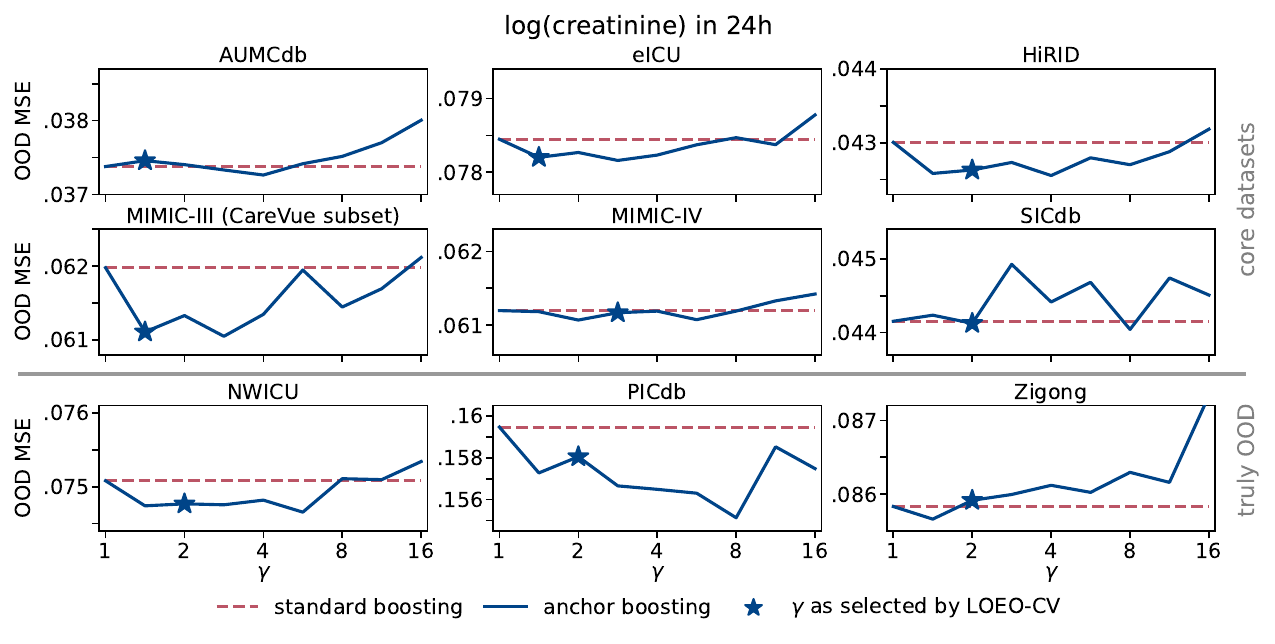}
    \caption{
        \label{fig:crea_algbm}
        Anchor boosting's OOD MSE predicting log(creatinine) as a function of $\gamma$, using one-hot-encoded dataset ID as anchor. See \cref{subsec:gamma} for details on the LOEO-CV model selection.
    }
\end{figure}

\Cref{fig:crea_algbm} shows anchor boosting's OOD mean squared error (MSE) predicting log(crea\-tinine) in 24 hours.
For the four targets eICU, HiRID, MIMIC-III, and NWICU, anchor regularization with $\gamma>1$ yields a considerable improvement of around 1\% of MSE.
For the truly OOD pediatric intensive care center PICdb, anchor regularization with $\gamma > 1$ yields a large improvement of around 3\% MSE.
Such small percentage improvements are substantial, as we discuss in \cref{sec:results:rescaling}.

\begin{figure}[tbph]
    \includegraphics[width=\textwidth]{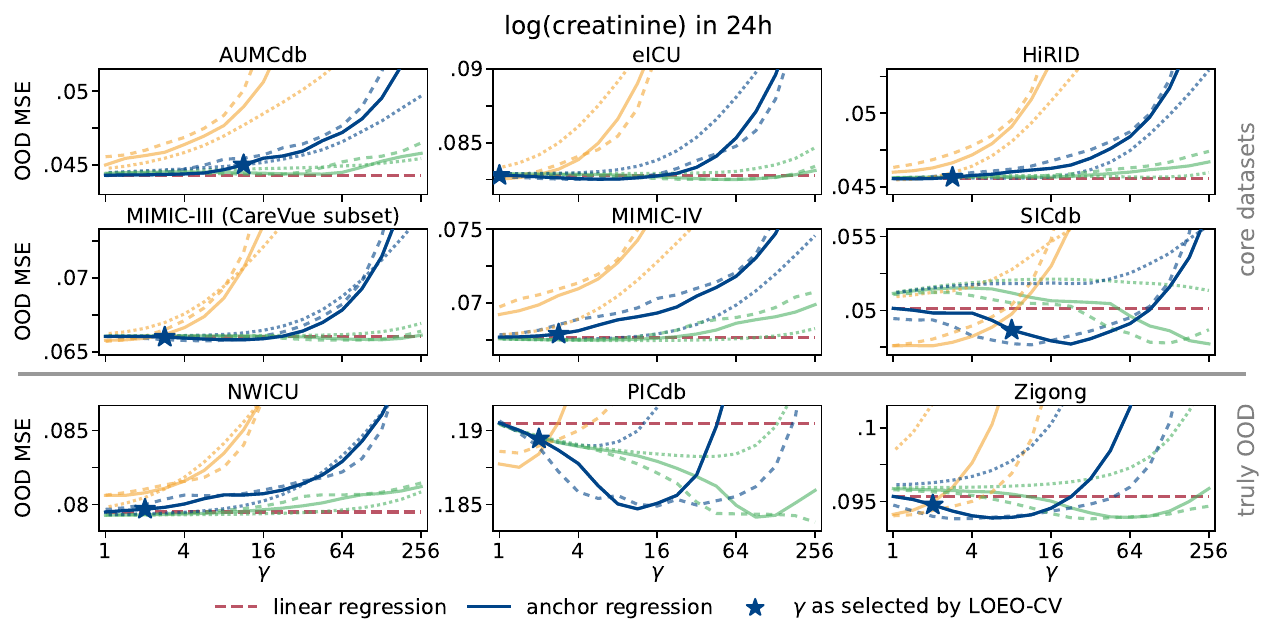}
    \caption{
        \label{fig:crea_anchor_colored}
        Linear anchor regression's OOD MSE predicting log(creatinine) in 24 hours as a function of $\gamma$.
        We add an elastic-net regularization term $\lambda \left( \eta \|\beta\|_1 + (1 - \eta) \| \beta \|_2^2 \right)$ to \cref{eq:anchor,eq:anchor_general}.
        Performances are colored by $\lambda = \lambda_\mathrm{max} / 10^2$ (orange), $\lambda_\mathrm{max} / 10^3$ (blue), and $\lambda_\mathrm{max} / 10^4$ (green). %
        Lasso ($\eta=1$) is dashed, elastic net ($\eta = 0.5)$ solid, and ridge ($\eta=0$) dotted.
    }
\end{figure}

\Cref{fig:crea_anchor_colored} shows the OOD MSE of linear anchor regression for the same task.
Anchor regularization with $\gamma > 1$ yields considerable improvements of around 1\% -- 3\% for the targets SICdb, Zigong, and PICdb.
It also leads to apparently minor improvements for eICU and MIMIC-III, which we show to be substantial in \cref{sec:results:rescaling}.

\Cref{fig:kidney_algbm_which_anchor} in appendix \ref{app:additional_figures} shows anchor boosting's OOD area under the precision-recall curve (AuPRC) predicting acute kidney injury within 48 hours, the classification task corresponding to log(creatinine) regression.
AuPRC is a common metric for predicting rare events on ICU data.
Anchor regularization with $\gamma > 1$ yields a considerable improvement of approximately 1\% of AuPRC for a subset of targets and a large improvement of around 4\% of AuPRC for the truly OOD target PICdb.

We show the OOD performance of anchor regression and boosting performance when predicting log(lactate) in 4 hours and circulatory failure in 8 hours in \cref{fig:circ_algbm_tune,fig:lact_anchor_which_anchor,fig:lact_algbm_which_anchor}.
Anchor regularization with $\gamma>1$ tends to improve performance for some of the target domains, often those which we pre-specified as ``truly OOD'' in \cref{sec:data:heterogeneity}.

\subsection{The performance gains are largest for the most OOD domains}
\label{sec:results:rescaling}
In their theory, \citet{rothenhausler2021anchor} assume that differences between environments are induced by shifts that are linear in the anchor variables.
They then show that anchor regression minimizes the worst-case error over distributions generated by shifts in the same direction as the shifts in the training data, but of larger magnitude.
The linear shift assumption implies that the residual noise levels, or task difficulties, are the same between environments.
Consequently, the largest errors occur exactly for environments with large shifts, that is, environments that are most out-of-distribution.

In \cref{sec:results:1}, we observe that anchor regularization improves performance for certain target domains.
However, the variance of MSEs and AuPRCs between different domains, both OOD and in-distribution, is much larger than the scale of improvement through anchor regularization.
That is, the assumption of constant noise level between environments does not apply and the environments with the largest OOD error are not necessarily those with the largest shift.

To verify whether anchor regularization improves performance for the domains with the largest shift, we rescale by asking: If we train a model on target domain data only, how many patients do we need to match the anchor model's OOD performances on that target domain?

To answer this, for 20 seeds, we draw increasing subsets of sizes 25, 35, 50, $\ldots$ of patient IDs from the target domain's train set and fit a model on data from these patients only.
We select hyperparameters using 5-fold cross-validation on the subsampled patients.
The median performances of these models, calculated over the 20 seeds, improves as more target domain patients are available.

\begin{figure}[tbph]
    \centering
    \hspace{1cm}
    \includegraphics[width=0.9\textwidth]{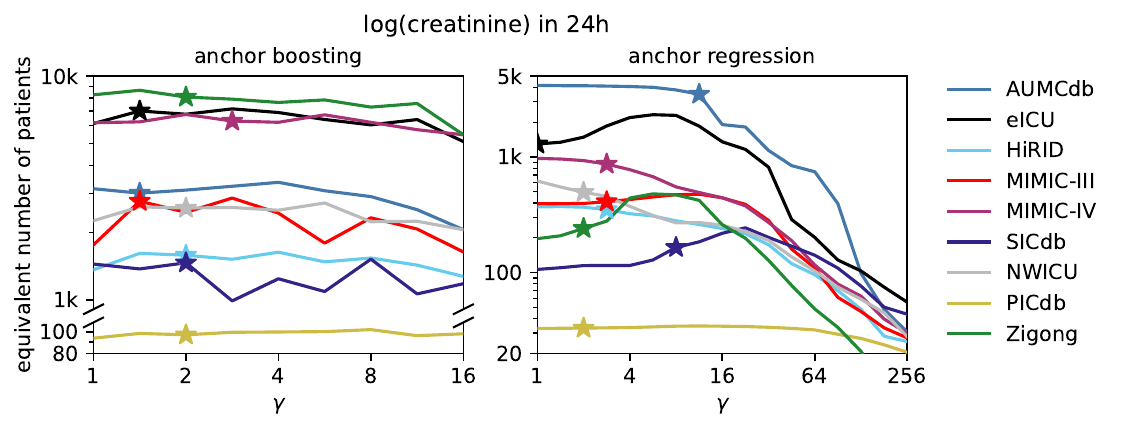}
    \caption{
        \label{fig:crea_rescaled}
        Differently expressed OOD performances predicting log(creatinine) in 24 hours as a function of $\gamma$.
        The performance on the y-axis is
        the number of patients from the target domain required to match
        nonlinear anchor boosting's (left) and linear anchor regression's (right) OOD performance.
    }
\end{figure}

In \cref{fig:crea_rescaled}, for each anchor model and $\gamma > 1$, we display the minimal number of patient IDs necessary for this median performance to match the anchor model's OOD performance.
We linearly interpolate $\log(\mathrm{num. patients} ) \sim \mathrm{performance}$ in between patient numbers.
We use this value of required patients to measure dissimilarity: The fewer patients needed from the target domain, the less value the large external dataset has for that task on the target, and thus the more dissimilar the target domain is.
We expect anchor regression to improve performance for the target domains with the largest shifts, effectively lifting the minimal number of patients required to match performance.

For both linear and boosted models, the target PICdb is clearly the most OOD, with less than 100 patients required to match the anchor models' performance.
However, while for PICdb the possible improvements are the largest, both in relative and in absolute terms, the improvement is relatively small after rescaling.
In this new scale, the improvement of anchor boosting for MIMIC-III and anchor regression for eICU, SICdb, and Zigong are the most impressive, with around twice as many patients from the target domain required to match the possible improvement of the anchor method.
For both eICU and MIMIC-III, the possible improvements of anchor regression appear minor in \cref{fig:crea_anchor_colored}, but reveal to be substantial when expressed in terms of equivalent patient samples
in \cref{fig:crea_rescaled}.
Finally, the greatest improvements of anchor regularization appear for the domains where the fewest in-distribution patient samples are required to match the performance of the model trained on external data.
That is, anchor boosting and linear anchor regression improves generalization to the domains that are most OOD.

We present equivalent figures for the remaining tasks in \cref{fig:lact_rescaled,fig:classification_rescaled} in appendix \ref{app:additional_figures}.
The results are similar.

\subsection{Selecting \texorpdfstring{$\gamma$}{gamma} is difficult}\label{subsec:gamma}
In \cref{fig:circ_algbm_tune,fig:crea_rescaled,fig:crea_anchor_colored,fig:lact_algbm_which_anchor,fig:lact_anchor_which_anchor} and \cref{fig:crea_algbm_tune,fig:lact_algbm_tune,fig:circ_algbm_which_anchor,fig:crea_anchor_which_anchor,fig:crea_algbm_which_anchor,fig:kidney_algbm_which_anchor,fig:lact_anchor_colored,fig:lact_rescaled,fig:classification_rescaled} in appendix \ref{app:additional_figures}, we mark the value of $\gamma$, selected for the default choice of hyperparameters by leave-one-environment-out cross-validation (LOEO-CV), minimizing the average OOD MSE for regression and the average OOD negative log-likelihood for classification, with a star.
We prefer the negative log-likelihood over AuPRC for model selection.
As a proper scoring rule, the log-likelihood is sensitive to calibration and thus better captures overfitting than rank-based metrics \citep{Friedman2000Additive}.

Linear anchor regression's robustness guarantees \citep{rothenhausler2021anchor} guard worst-case performance over a set of distribution shifts similar to those found in the training data, but with a magnitude scaled by $\gamma$.
Thus, the tuning parameter $\gamma$ should be chosen proportional to the expected strength of perturbations for the new target, relative to the perturbations seen in the training data.
If the shifts between the core datasets' distributions are similar enough, one would expect LOEO-CV to select a value of $\gamma$ close to 1.
In particular, one could argue that LOEO-CV on the core datasets is not a good tool to select $\gamma$ for the truly OOD datasets, which we expect to be more dissimilar.
In practice, LOEO-CV typically selects a value of $\gamma \in [1, 4]$, a reasonable choice for the core datasets, but 
too small for the truly OOD datasets.

Anchor regression aims to
improve worst-case, not average-case performance
to new environments. Thus, one could argue to minimize the worst-case (instead of average) MSE or negative log-likelihood over the OOD targets with LOEO-CV.
However, as discussed in \cref{sec:results:rescaling}, varying noise levels between domains make a comparison between them difficult.

Finally, \citet{rothenhausler2021anchor} suggest to select linear anchor regression's $\gamma$ based on prior knowledge about shift sizes, as an alternative to LOEO-CV.
In \cref{sec:results:hyperparameters}, we observe an interaction between conventional regularization and the optimal value for $\gamma$ of linear anchor regression, making such an approach problematic.

The interpretation of $\gamma$ as the relative shift magnitude in the test domain is theoretically grounded for linear anchor regression.
While we intuitively expect a similar relationship in the nonlinear case, there are no formal guarantees.
Empirically, $\gamma$ serves as a parameter to achieve best worst-case out-of-distribution performance, making LOEO-CV a practical tool for model selection.

\subsection{On the choice of other hyperparameters}
\label{sec:results:hyperparameters}
As mentioned in \cref{sec:methods:anchor_regression}, we add an elastic net regularization term of the form $\lambda \left( \eta \|\beta\|_1 + (1 - \eta) \| \beta\|_2^2 \right)$ to \cref{eq:anchor,eq:anchor_general} for linear anchor regression. 
Let $\lambda_\mathrm{max}$ be minimal such that all parameters other than the intercept of a lasso model ($\eta=\gamma=1$) with $\lambda=\lambda_\mathrm{max}$ are zero.
In \cref{fig:crea_anchor_colored}, we show the performance of anchor regression predicting log(creatinine) in 24 hours with $\eta = 0, 0.5, 1$ and $\lambda = 10^{-2} \lambda_\mathrm{max}, 10^{-3} \lambda_\mathrm{max}, 10^{-4} \lambda_\mathrm{max}$.
We observe an interaction between $\gamma$ and $\lambda$:
As we increase the amount of conventional regularization $\lambda$ by a factor of 10, the optimal amount of anchor regularization $\gamma$ decreases by around 10.
We observe the same effect when predicting log(lactate) in 4 hours, see \cref{fig:lact_anchor_colored} in appendix \ref{app:additional_figures}.

We are not aware of any prior work that observed or explained this phenomenon.
\citet{kostin2024achievable} explore the interaction of anchor and ridge regularization, but none of their results suggests the relationship $\lambda \cdot \gamma_\mathrm{optimal} = \mathrm{const}$, where $\gamma_\mathrm{optimal}$ optimizes the target domain performance for fixed conventional regularization $\lambda$.

\begin{figure}[ht]
    \includegraphics[width=\textwidth]{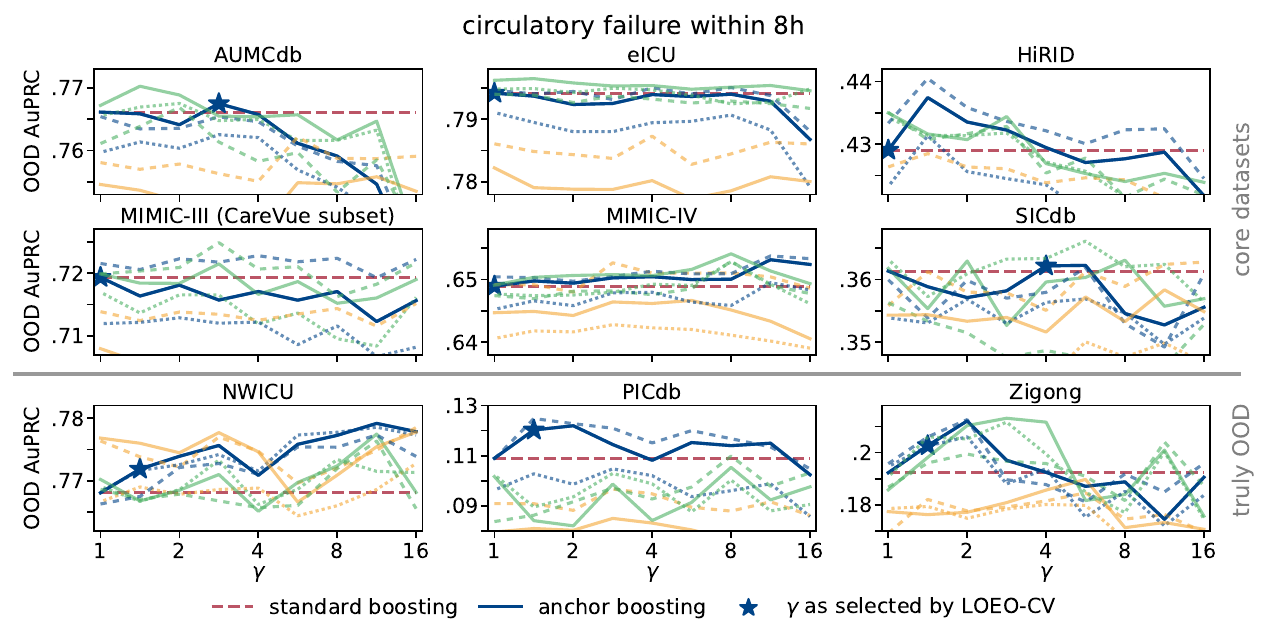}
    \caption{
        \label{fig:circ_algbm_tune}
        Boosted anchor classification's OOD AuPRC (larger is better) predicting circulatory failure within 8 hours as a function of $\gamma$.
        We vary the number of trees from 500 (dotted), 1000 (solid), to 2000 (dashed) and the trees' maximal depth from 2 (orange), 3 (blue), to 4 (green).
    }
\end{figure}

Due to the algorithm's increased variance, the corresponding plots for nonlinear anchor boosting are more difficult to interpret.
In \cref{fig:circ_algbm_tune}, we show anchor boosting's AuPRC predicting circulatory failure within 8 hours.
We vary the number of trees from 500, 1000, to 2000, and the trees' maximal depth from 2, 3, to 4.
We do not observe a clear effect that a lower value of $\gamma$ being more optimal for a more strongly conventionally regularized boosted model.
In contrast, 
for a fixed maximal depth, the shape of the models' AuPRC with a varying number of trees stays similar.
This suggests that anchor regularization has mainly an effect in the early boosting iterations.
We observe the same behavior when predicting log(lactate), log(creatinine), and acute kidney injury, see \cref{fig:lact_algbm_tune,fig:crea_algbm_tune} in appendix \ref{app:additional_figures}.

For both linear anchor regression and nonlinear anchor boosting models, we observe that conventional regularization is important.
Anchor regularization is an additional tool to the existing toolbox of modeling choices.

\subsection{Which variables are good choices to use as anchors?}
\label{sec:results:which_anchor}
So far, we have used a one-hot encoding of the discrete dataset ID as an anchor.
This is a plausible choice since the anchor variables should describe exogenous variables acting on the system and as anchor regression improves robustness in the shift directions induced by the anchor variable.
As we are interested in generalizing to new datasets, this suggests using the dataset ID as anchor.

Some of the ICU datasets we consider include additional variables encoding heterogeneity: (i) the year of admission, (ii) an identifier for the hospital ward a patient was assigned to, (iii) a patients insurance type (for example, private or public), (iv) the type of admission (surgical, medical or other), (v) a patient's ID, and (vi) ICD9 or ICD10 codes of diagnoses.
We summarize the availability of these variables in the core datasets in \cref{tab:other_anchors}.

\begin{table}[ht]
\center
\caption{
    \label{tab:other_anchors}
    Number of categories (unique entries) of possible anchors encoding heterogeneity in the core datasets' 85\% training set.
}
\begin{tabular}{lcccccc}
dataset & admission  & year & wards & insurance & ICD codes & patients\\
\midrule
AUMCdb & 4 & 2 & 3 & - & - & 16'958 \\
eICU & 4 & 2 & 331 & - & 135 & 159'812\\
HiRID & 3 & - & - & - & - & 28'479\\
MIMIC-III (CV) & 4 & - & 13 & 5 & 207 & 23'191 \\
MIMIC-IV & 4 & 16 & 6 & 5 & 203 & 55'237 \\
SICdb & 3 & 9 & 4 & - & 160 & 18'184 \\
\end{tabular}

\end{table}

One can imagine that these variables encode heterogeneity similar to some of the heterogeneity between data sets described in \cref{sec:data:heterogeneity}.
Excluding ICD codes, they are measured at the beginning of the ICU stay, and thus potentially satisfy exogeneity.
ICD codes are classifiers for patient diagnoses and are typically assigned at the end of a patient's stay.
They are thus endogenous, violating a main assumption of anchor regression.

\begin{figure}[tbph]
    \includegraphics[width=\textwidth]{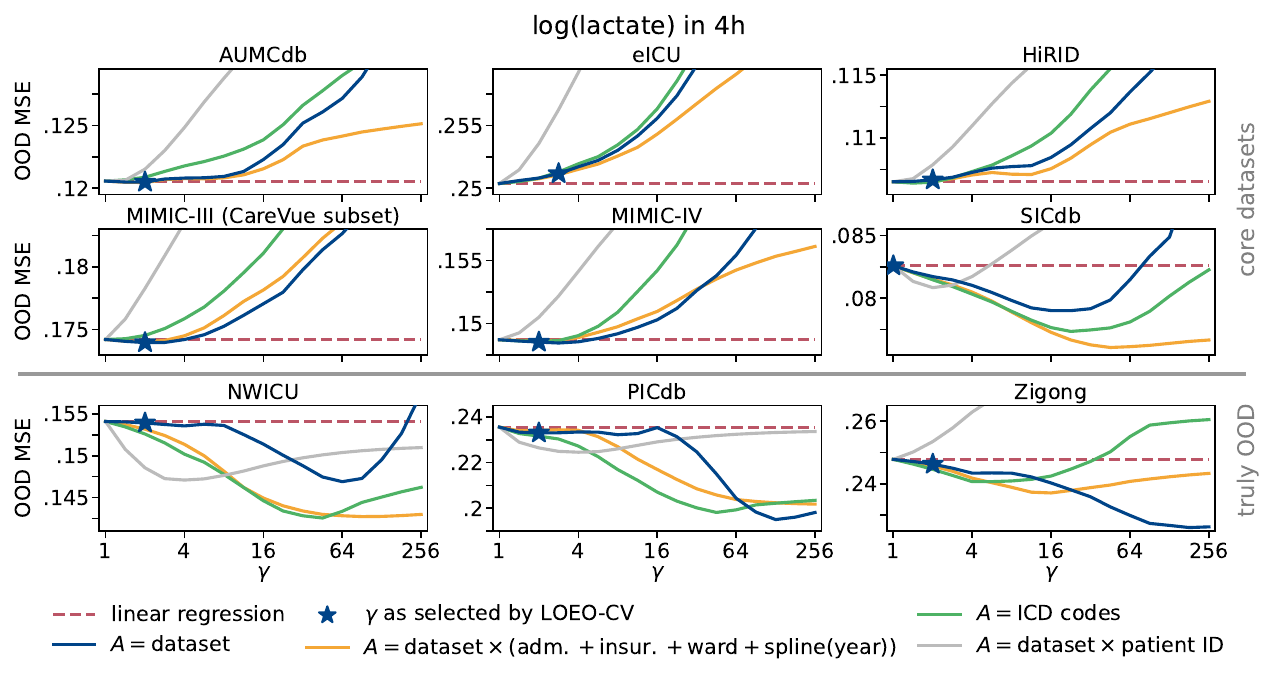}
    \caption{
        \label{fig:lact_anchor_which_anchor}
        Linear anchor regression's OOD MSE predicting log(lactate) in 4 hours as a function of $\gamma$ and the anchor used.
    }
\end{figure}

\begin{figure}[tbph]
    \includegraphics[width=\textwidth]{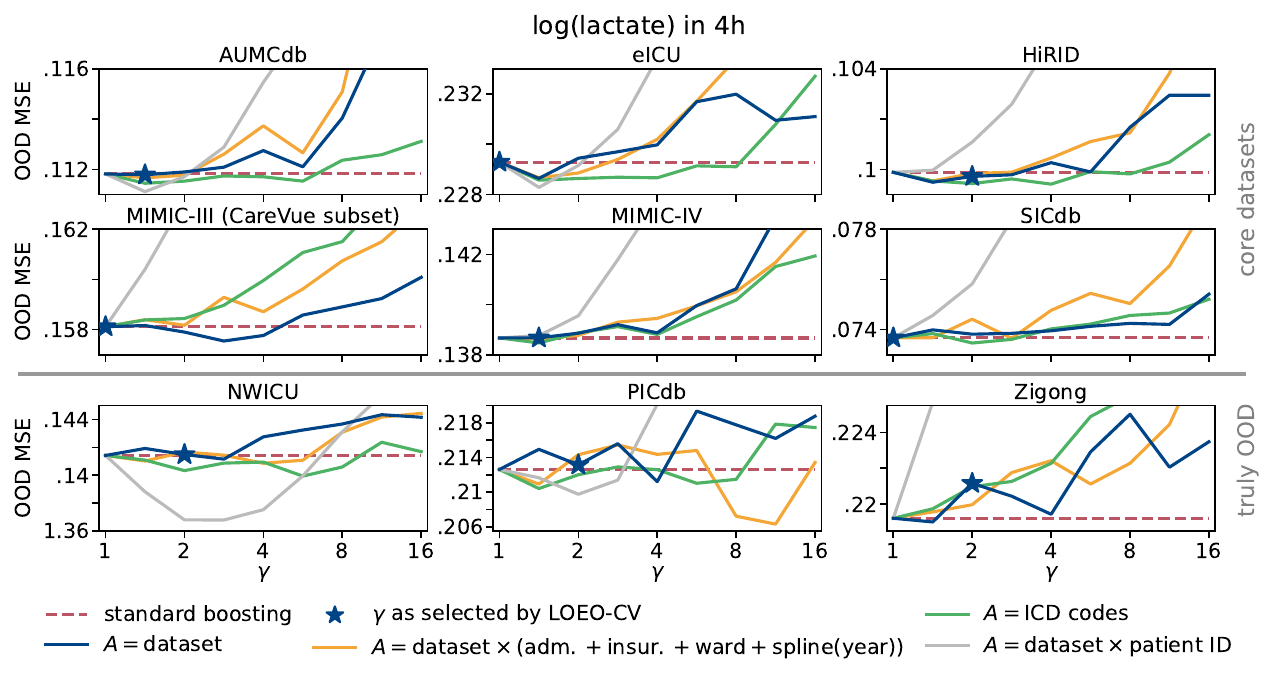}    \caption{
        \label{fig:lact_algbm_which_anchor}
        Boosted anchor regression's OOD MSE predicting log(lactate) in 4 hours as a function of $\gamma$ and the anchor used.
    }
\end{figure}

In \cref{fig:lact_anchor_which_anchor,fig:lact_algbm_which_anchor} we show linear anchor regression and nonlinear anchor boosting's MSE for predicting log(lactate) in 4 hours as a function of $\gamma$ and the anchor used.
We compare performances when (i)~using the dataset ID, (ii)~interacting the dataset ID with a sum of the one-hot encodings of admission type, insurance type, and hospital ward, and a four-knot spline over years, (iii)~a multiple-hot encoding of the ICD codes, and (iv) the patient IDs, as anchors.

For every choice of anchors we tested, linear anchor regression with $\gamma > 1$ potentially improves OOD MSE on SICdb and the truly OOD targets NWICU and PICdb.
Nonlinear anchor boosting shows a similar, albeit noisier, behavior.

Particularly interesting is that anchor regularization still improves performance when the endogenous ICD codes are used as anchors.
This suggests that anchor regression is robust to some violations of the anchor exogeneity assumption, an important consideration for practitioners without access to purely exogenous variables.
This general effectiveness, both with endogenous anchors and with anchors unavailable in the target domain, points to a wider and more flexible applicability of anchor regression than its theory might suggest.

Results are similar when predicting log(creatinine), acute kidney injury, and circulatory failure, see \cref{fig:crea_anchor_which_anchor,fig:crea_algbm_which_anchor,fig:kidney_algbm_which_anchor,fig:circ_algbm_which_anchor} in appendix \ref{app:additional_figures}.

\subsection{Refitting models using few samples from the target}
\label{sec:results:refitting}
Domain generalization is difficult in our application, and the observed prediction performances are rather poor.
In practice, it would be highly beneficial to have few samples from the target distribution available to refit models (domain adaptation).

\begin{figure}[tbph]
    \includegraphics[width=\textwidth]{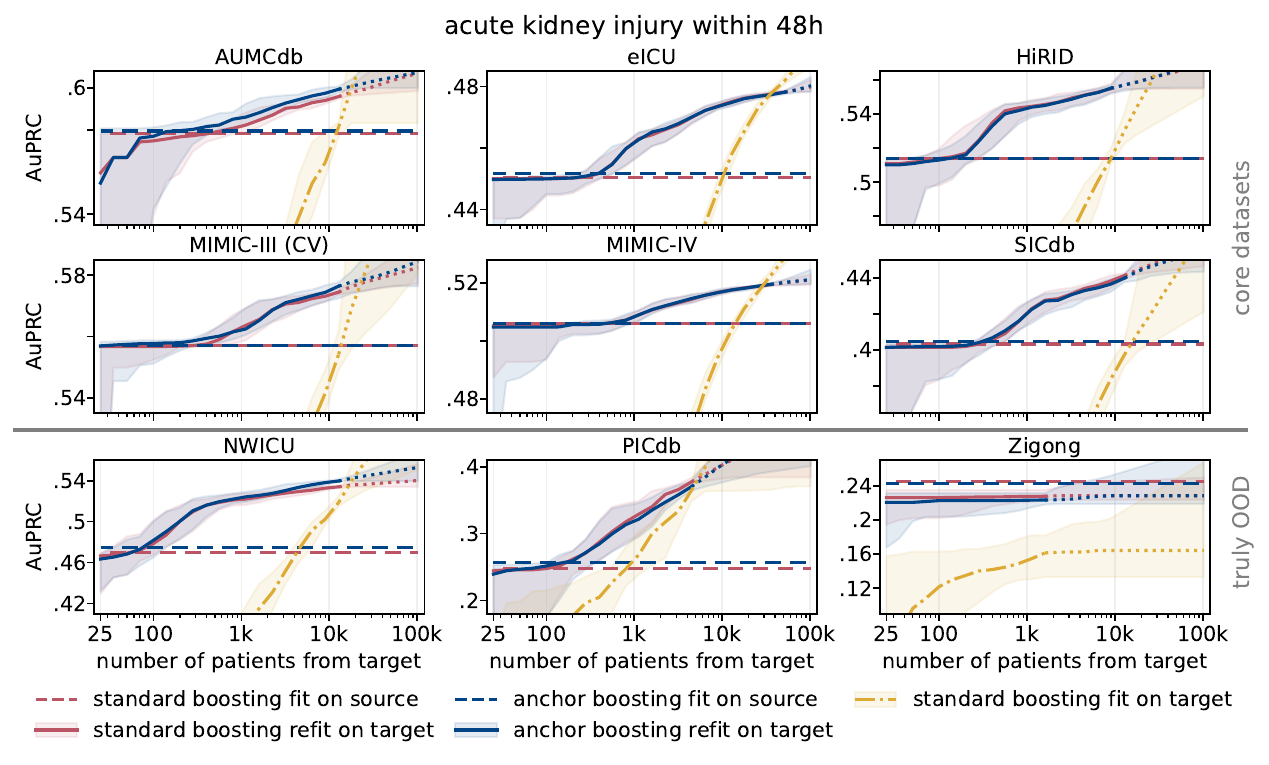}
    \caption{
        \label{fig:kidney_algbm_n_samples}
        AuPRC (larger is better) predicting acute kidney injury within 48 hours as a function of available patients from the target dataset.
        Lines are medians and shaded areas are 80\% credible sets over 20 different subsampling seeds.
    }
\end{figure}

We apply the refitting methodology of \cref{sec:methods:refitting,sec:methods:refitting_nonlinear} to the ICU data.
For 20 seeds, we draw increasing subsets of sizes 25, 35, 50, 70, 100, ... of patient IDs from the target's train set and then refit the anchor models on these patients.
We select $\gamma \in \{1, 2, 4, \ldots \}$, the decay rate $\mathrm{dr} \in \{0, 0.2, \ldots, 1\}$, and the prior's width $\alpha \in \{10, \sqrt{10}, \ldots, 10^{-3}\}$ with 5-fold cross-validation on the sampled target data.
We also fit models using only the sampled target data.
For each seed, we fit monotonous cubic splines to the model performances and use these to extrapolate beyond the number of patient samples available for each domain.
We show the performance of these approaches predicting acute kidney injury within 48 hours in \cref{fig:kidney_algbm_n_samples} and predicting log(creatinine), log(lactate), and circulatory failure in \cref{fig:crea_algbm_n_samples,fig:lact_algbm_n_samples,fig:circ_algbm_n_samples,fig:crea_anchor_n_samples,fig:lact_anchor_n_samples,fig:circ_glm_n_samples,fig:kidney_glm_n_samples} in appendix \ref{app:additional_figures}.

We can clearly see the three regimes as described in \cref{sec:methods:regimes}.

\subsection{Three regimes and information in external data}
\label{sec:results:regimes}
\begin{figure}[tbph]
    \centering
    \includegraphics[width=0.75\textwidth]{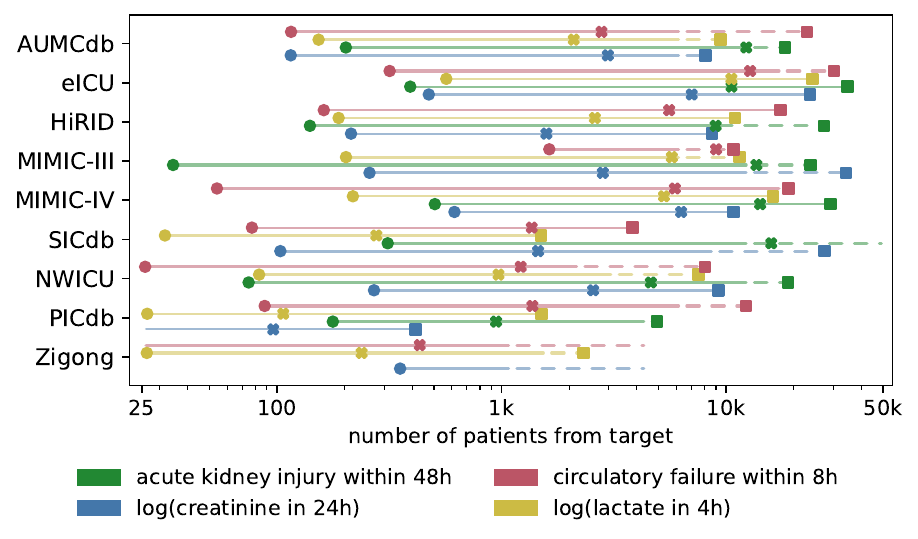}
    \caption{
        \label{fig:regimes}
        Regime transitions for boosted tree models as described in \cref{sec:methods:regimes} and \cref{fig:illustration}.
        \circlemarker~denotes the regime transition i $\to$ ii,
        \squaremarker~the regime transition ii $\to$ iii, and
        \xmarker~denotes the external data's value.
    }
\end{figure}

We described the three regimes and their transition points in \cref{sec:methods:regimes}.
We summarize the empirical transition points for anchor boosting applied to acute kidney injury, circulatory failure, log(creatinine), and log(lactate) prediction in \cref{fig:regimes}.
We believe that these results are important to understand the potential of using external data for domain adaptation.
We show the equivalent plot for linear anchor and logistic regression in \cref{fig:regimes_linear} in appendix \ref{app:additional_figures}.

The regime transitions and the value of external data vary by dataset and task.
We observe that SICdb appears to be the most dissimilar amongst the core datasets, comparable to NWICU.
For the remaining five core datasets, it appears that the external data is worth around 1'500--15'000 in-distribution patients' data. 
Here, the external data has no additional value once around 10'000--50'000 in-distribution patient's data is available.

For SICdb and the truly OOD datasets the results are more extreme.
The external data is only worth 100 in-distribution patients' data when predicting log(lactate) or log(creatinine) on PICdb.

It is interesting to compare this to the ``patient equivalent'' values reported by \citet{burger2025foundation}.
Using a large foundation model based on scaling, they find similar values, reporting domain generalization performance worth several thousand local patients.

\section{Discussion}
We address the challenge of domain generalization in multi-center ICU predictive models using principles of causal invariance.
We introduce anchor boosting, a novel nonlinear extension of anchor regression.
As a rare finding of success in ICU data domain generalization, both linear and nonlinear anchor methods frequently improve out-of-distribution performance, particularly for the most out-of-distribution target domains.

We also propose a framework that quantifies the value of external out-of-distribution data.
The framework quantifies the value of external data by determining the number of target-domain samples required to match an OOD model's performance.
It uses this to identify three distinct regimes of data utility: domain generalization, domain adaptation, and target-only training.
This taxonomy provides a data-driven methodology for practitioners to decide how to integrate external data.
Our results show that while external data is valuable when %
target data is scarce, its value diminishes and it eventually becomes obsolete as more in-distribution data becomes available.

Some challenges remain.
The empirical results are not uniformly positive across all settings.
Furthermore, the selection of the anchor regularization parameter $\gamma$ is non-trivial and interacts with conventional regularization.
Finally, the choice of the anchor variable remains subjective and a more data-driven methodology is missing.

However, given the inherent difficulty of domain generalization in the multi-center ICU setting, the proposed framework offers a path forward.
Using a distributionally robust prediction from external multi-source data as a prior, towards which we shrink an algorithm based on relatively few target data, proves to be a coherent and promising approach.

Finally, our proposed framework for evaluating external training data utility is highly general and we would welcome its application to other (multi-source) domain adaptation learning problems.

\section*{Data and code availability}
Our study uses harmonized data and resources expanding the \href{https://github.com/eth-mds/ricu/}{\texttt{ricu}} R-package \citep{bennett2023ricu} from prior work by \citet{burger2024foundationmodelscriticalcare,burger2025foundation}, who describe a large-scale multi-center ICU dataset.
This extended version of \texttt{ricu} includes additional code and configuration files to harmonize data from additional data sources (NWICU, PICdb, and Zigong) and to extract additional variables and will be published together with \citet{burger2025foundation}.
See \hyperlink{github.com/ratschlab/icarefm}{\texttt{https://github.com/ratschlab/icarefm}} for details.

The raw datasets eICU, HiRID, MIMIC-III, MIMIC-IV, SICdb, NWICU, PICdb, and Zigong are available on PhysioNet \citep{goldberger2000physiobank} upon completion of CITI's ``Data or Specimens Only Research'' course.
HiRID and SICdb require additional approval from the dataset owners.
AUMCdb is not on PhysioNet and requires a separate access request.
If legally feasible, the harmonized multi-center ICU dataset will be published as part of \citet{burger2025foundation}.

The code to compute outcomes and features can be found in the GitHub repository \hyperlink{https://github.com/eth-mds/icu-features}{\texttt{github.com/eth-mds/icu-features}}.
The code to create figures and tables in this manuscript can be found at \hyperlink{https://github.com/mlondschien/icu-benchmarks}{\texttt{github.com/mlondschien/icu-benchmarks}}.

We implement anchor boosting in the \texttt{anchorboosting} software package for Python. %
See the GitHub repository at \href{https://github.com/mlondschien/anchorboosting/}{\texttt{github.com/mlondschien/anchorboosting}} and the documentation at \href{https://anchorboosting.readthedocs.io/}{\texttt{anchorboosting.readthedocs.io}} for more details.
We use the \href{https://ivmodels.readthedocs.io/en/latest/api/ivmodels.models.html#ivmodels.models.anchor_regression.AnchorRegression}{Anchor Regression} implementation of the \href{https://github.com/mlondschien/ivmodels/}{\texttt{ivmodels}} software package for Python \citep{londschien2024weak}.

\section*{Funding}
Malte Londschien is partially supported by the ETH Foundations of Data Science.

\section*{Conflicts of interest}
The authors declare no conflicts of interest.

\section*{Acknowledgements}
We gratefully acknowledge early access to the harmonized dataset and \texttt{ricu} resources provided by \citet{burger2024foundationmodelscriticalcare,burger2025foundation}.
We would like to thank Cyrill Scheidegger,  Dinara Veshchezerova, Dominik Rothenhäusler, Gianna Wolfisberg, Hugo Y\`eche, Jonas Peters,  Julia Kostin, Markus Ulmer,  Maybritt Schillinger, Michael Law, Nicolas Bennet, Niklas Pfister, Olga Mineeva, Patrick Rockenschaub, Paola Malsot, Sorawit Saengkyongam, and  Yuansi Chen for helpful discussions and comments. Finally, we thank the editor, associate editor and reviewers for their constructive comments.

\bibliography{bib}

\begin{thebibliography}{}

\bibitem[\protect\citeauthoryear{{Acute Kidney Injury Work Group}}{{Acute
  Kidney Injury Work Group}}{2012}]{kidney2012kdigo}
{Acute Kidney Injury Work Group} (2012).
\newblock {KDIGO} clinical practice guideline for acute kidney injury; section
  2: {AKI} definition.
\newblock {\em Kidney International Supplements\/}~{\em 2}, 19--36.


\bibitem[\protect\citeauthoryear{Arjovsky, Bottou, Gulrajani, and
  Lopez-Paz}{Arjovsky et~al.}{2019}]{arjovski2019invariant}
Arjovsky, M., L.~Bottou, I.~Gulrajani, and D.~Lopez-Paz (2019).
\newblock Invariant risk minimization.
\newblock arXiv:1907.02893.

\bibitem[\protect\citeauthoryear{Barak-Corren, Chaudhari, Perniciaro, Waltzman,
  Fine, and Reis}{Barak-Corren et~al.}{2021}]{barak2021prediction}
Barak-Corren, Y., P.~Chaudhari, J.~Perniciaro, M.~Waltzman, A.~M. Fine, and
  B.~Y. Reis (2021).
\newblock Prediction across healthcare settings: a case study in predicting
  emergency department disposition.
\newblock {\em npj Digital Medicine\/}~{\em 4\/}(1), 169.


\bibitem[\protect\citeauthoryear{Bennett, Ple{\v{c}}ko, Ukor, Meinshausen, and
  B{\"u}hlmann}{Bennett et~al.}{2023}]{bennett2023ricu}
Bennett, N., D.~Ple{\v{c}}ko, I.-F. Ukor, N.~Meinshausen, and P.~B{\"u}hlmann
  (2023).
\newblock ricu: R’s interface to intensive care data.
\newblock {\em GigaScience\/}~{\em 12}, giad041.


\bibitem[\protect\citeauthoryear{Brown, Mann, Ryder, Subbiah, Kaplan, Dhariwal,
  Neelakantan, Shyam, Sastry, Askell, et~al.}{Brown
  et~al.}{2020}]{brown2020language}
Brown, T., B.~Mann, N.~Ryder, M.~Subbiah, J.~D. Kaplan, P.~Dhariwal,
  A.~Neelakantan, P.~Shyam, G.~Sastry, A.~Askell, et~al. (2020).
\newblock Language models are few-shot learners.
\newblock {\em Advances in neural information processing systems\/}~{\em 33},
  1877--1901.


\bibitem[\protect\citeauthoryear{B{\"u}hlmann}{B{\"u}hlmann}{2020}]{buhlmann2020invariance}
B{\"u}hlmann, P. (2020).
\newblock Invariance, causality and robustness.
\newblock {\em Statistical Science\/}~{\em 35\/}(3), 404--426.


\bibitem[\protect\citeauthoryear{Burger, Chopard, Londschien, Sergeev,
  Y{\`e}che, Kuznetsova, Faltys, Gerdes, Leshetkina, B{\"u}hlmann, and
  R\"atsch}{Burger et~al.}{2025}]{burger2025foundation}
Burger, M., D.~Chopard, M.~Londschien, F.~Sergeev, H.~Y{\`e}che, R.~Kuznetsova,
  M.~Faltys, E.~Gerdes, P.~Leshetkina, P.~B{\"u}hlmann, and G.~R\"atsch (2025).
\newblock A foundation model for intensive care unlocking generalization across
  tasks and domains at scale.
\newblock medRxiv:2025.07.25.25331635.

\bibitem[\protect\citeauthoryear{Burger, Sergeev, Londschien, Chopard,
  Y{\`e}che, Gerdes, Leshetkina, Morgenroth, Bab{\"u}r, Bogojeska, Faltys,
  Kuznetsova, and R{\"a}tsch}{Burger
  et~al.}{2024}]{burger2024foundationmodelscriticalcare}
Burger, M., F.~Sergeev, M.~Londschien, D.~Chopard, H.~Y{\`e}che, E.~Gerdes,
  P.~Leshetkina, A.~Morgenroth, Z.~Bab{\"u}r, J.~Bogojeska, M.~Faltys,
  R.~Kuznetsova, and G.~R{\"a}tsch (2024).
\newblock Towards foundation models for critical care time series.
\newblock arXiv:2411.16346.

\bibitem[\protect\citeauthoryear{Christiansen, Pfister, Jakobsen, Gnecco, and
  Peters}{Christiansen et~al.}{2022}]{Christiansen2022Causal}
Christiansen, R., N.~Pfister, M.~E. Jakobsen, N.~Gnecco, and J.~Peters (2022).
\newblock A causal framework for distribution generalization.
\newblock {\em IEEE Transactions on Pattern Analysis and Machine
  Intelligence\/}~{\em 44\/}(10), 6614--6630.


\bibitem[\protect\citeauthoryear{Desautels, Calvert, Hoffman, Mao, Jay,
  Fletcher, Barton, Chettipally, Kerem, and Das}{Desautels
  et~al.}{2017}]{desautels2017using}
Desautels, T., J.~Calvert, J.~Hoffman, Q.~Mao, M.~Jay, G.~Fletcher, C.~Barton,
  U.~Chettipally, Y.~Kerem, and R.~Das (2017).
\newblock Using transfer learning for improved mortality prediction in a
  data-scarce hospital setting.
\newblock {\em Biomedical Informatics Insights\/}~{\em 9}, 1178222617712994.


\bibitem[\protect\citeauthoryear{Friedman, Hastie, and Tibshirani}{Friedman
  et~al.}{2000}]{Friedman2000Additive}
Friedman, J., T.~Hastie, and R.~Tibshirani (2000).
\newblock {Additive logistic regression: a statistical view of boosting}.
\newblock {\em Annals of Statistics\/}~{\em 28\/}(2), 337--407.


\bibitem[\protect\citeauthoryear{Friedman}{Friedman}{2001}]{friedman2001greedy}
Friedman, J.~H. (2001).
\newblock Greedy function approximation: a gradient boosting machine.
\newblock {\em Annals of statistics\/}~{\em 29\/}(5), 1189--1232.


\bibitem[\protect\citeauthoryear{Goldberger, Amaral, Glass, Hausdorff, Ivanov,
  Mark, Mietus, Moody, Peng, and Stanley}{Goldberger
  et~al.}{2000}]{goldberger2000physiobank}
Goldberger, A.~L., L.~A. Amaral, L.~Glass, J.~M. Hausdorff, P.~C. Ivanov, R.~G.
  Mark, J.~E. Mietus, G.~B. Moody, C.-K. Peng, and H.~E. Stanley (2000).
\newblock {PhysioBank, PhysioToolkit, and PhysioNet: components of a new
  research resource for complex physiologic signals}.
\newblock {\em Circulation\/}~{\em 101\/}(23), e215--e220.


\bibitem[\protect\citeauthoryear{Gulrajani and Lopez-Paz}{Gulrajani and
  Lopez-Paz}{2021}]{gulrajani2020search}
Gulrajani, I. and D.~Lopez-Paz (2021).
\newblock In search of lost domain generalization.
\newblock In {\em International Conference on Learning Representations}.

\bibitem[\protect\citeauthoryear{Guo, Pfohl, Fries, Johnson, Posada,
  Aftandilian, Shah, and Sung}{Guo et~al.}{2022}]{guo2022evaluation}
Guo, L.~L., S.~R. Pfohl, J.~Fries, A.~E. Johnson, J.~Posada, C.~Aftandilian,
  N.~Shah, and L.~Sung (2022).
\newblock Evaluation of domain generalization and adaptation on improving model
  robustness to temporal dataset shift in clinical medicine.
\newblock {\em Scientific reports\/}~{\em 12\/}(1), 2726.


\bibitem[\protect\citeauthoryear{Guo, Pfohl, Fries, Posada, Fleming,
  Aftandilian, Shah, and Sung}{Guo et~al.}{2021}]{guo2021systematic}
Guo, L.~L., S.~R. Pfohl, J.~Fries, J.~Posada, S.~L. Fleming, C.~Aftandilian,
  N.~Shah, and L.~Sung (2021).
\newblock Systematic review of approaches to preserve machine learning
  performance in the presence of temporal dataset shift in clinical medicine.
\newblock {\em Applied clinical informatics\/}~{\em 12\/}(04), 808--815.


\bibitem[\protect\citeauthoryear{Hendrycks and Dietterich}{Hendrycks and
  Dietterich}{2019}]{hendrycks2019benchmarking}
Hendrycks, D. and T.~Dietterich (2019).
\newblock Benchmarking neural network robustness to common corruptions and
  perturbations.
\newblock In {\em International Conference on Learning Representations}.

\bibitem[\protect\citeauthoryear{H{\"u}ser, Lyu, Faltys, Pace, Hoche, Hyland,
  Y{\`e}che, Burger, Merz, and R{\"a}tsch}{H{\"u}ser
  et~al.}{2024}]{huser2024comprehensive}
H{\"u}ser, M., X.~Lyu, M.~Faltys, A.~Pace, M.~Hoche, S.~Hyland, H.~Y{\`e}che,
  M.~Burger, T.~M. Merz, and G.~R{\"a}tsch (2024).
\newblock A comprehensive ml-based respiratory monitoring system for
  physiological monitoring \& resource planning in the {ICU}.
\newblock medRxiv:2024.01.23.24301516.

\bibitem[\protect\citeauthoryear{Hyland, Faltys, H{\"u}ser, Lyu, Gumbsch,
  Esteban, Bock, Horn, Moor, Rieck, et~al.}{Hyland
  et~al.}{2020}]{hyland2020early}
Hyland, S.~L., M.~Faltys, M.~H{\"u}ser, X.~Lyu, T.~Gumbsch, C.~Esteban,
  C.~Bock, M.~Horn, M.~Moor, B.~Rieck, et~al. (2020).
\newblock Early prediction of circulatory failure in the intensive care unit
  using machine learning.
\newblock {\em Nature medicine\/}~{\em 26\/}(3), 364--373.


\bibitem[\protect\citeauthoryear{Johnson, Bulgarelli, Shen, Gayles, Shammout,
  Horng, Pollard, Hao, Moody, Gow, et~al.}{Johnson
  et~al.}{2023}]{johnson2023mimiciv}
Johnson, A.~E., L.~Bulgarelli, L.~Shen, A.~Gayles, A.~Shammout, S.~Horng, T.~J.
  Pollard, S.~Hao, B.~Moody, B.~Gow, et~al. (2023).
\newblock {MIMIC-IV}, a freely accessible electronic health record dataset.
\newblock {\em Scientific data\/}~{\em 10\/}(1), 1.


\bibitem[\protect\citeauthoryear{Johnson, Pollard, Shen, Lehman, Feng,
  Ghassemi, Moody, Szolovits, Anthony~Celi, and Mark}{Johnson
  et~al.}{2016}]{johnson2016mimiciiia}
Johnson, A.~E., T.~J. Pollard, L.~Shen, L.-w.~H. Lehman, M.~Feng, M.~Ghassemi,
  B.~Moody, P.~Szolovits, L.~Anthony~Celi, and R.~G. Mark (2016).
\newblock {MIMIC-III}, a freely accessible critical care database.
\newblock {\em Scientific data\/}~{\em 3\/}(1), 1--9.


\bibitem[\protect\citeauthoryear{Ke, Meng, Finley, Wang, Chen, Ma, Ye, and
  Liu}{Ke et~al.}{2017}]{ke2017lightgbm}
Ke, G., Q.~Meng, T.~Finley, T.~Wang, W.~Chen, W.~Ma, Q.~Ye, and T.-Y. Liu
  (2017).
\newblock {LightGBM}: A highly efficient gradient boosting decision tree.
\newblock {\em Advances in neural information processing systems\/}~{\em 30},
  3146--3154.


\bibitem[\protect\citeauthoryear{Kook, Sick, and B{\"u}hlmann}{Kook
  et~al.}{2022}]{kook2022distributional}
Kook, L., B.~Sick, and P.~B{\"u}hlmann (2022).
\newblock Distributional anchor regression.
\newblock {\em Statistics and Computing\/}~{\em 32\/}(3), 39.


\bibitem[\protect\citeauthoryear{Kostin, Gnecco, and Yang}{Kostin
  et~al.}{2024}]{kostin2024achievable}
Kostin, J., N.~Gnecco, and F.~Yang (2024).
\newblock Achievable distributional robustness when the robust risk is only
  partially identified.
\newblock {\em Advances in Neural Information Processing Systems\/}~{\em 37},
  83915--83950.


\bibitem[\protect\citeauthoryear{Li, Yang, Song, and Hospedales}{Li
  et~al.}{2017}]{li2017deeper}
Li, D., Y.~Yang, Y.-Z. Song, and T.~M. Hospedales (2017).
\newblock Deeper, broader and artier domain generalization.
\newblock In {\em Proceedings of the IEEE international conference on computer
  vision}, pp.\  5542--5550.

\bibitem[\protect\citeauthoryear{Li, Zeng, and Yu}{Li
  et~al.}{2019}]{li2019paediatric}
Li, H., X.~Zeng, and G.~Yu (2019).
\newblock Paediatric intensive care database (version 1.1.0).

\bibitem[\protect\citeauthoryear{Londschien and Bühlmann}{Londschien and
  Bühlmann}{2024}]{londschien2024weak}
Londschien, M. and P.~Bühlmann (2024).
\newblock Weak-instrument-robust subvector inference in instrumental variables
  regression: A subvector lagrange multiplier test and properties of subvector
  anderson-rubin confidence sets.
\newblock arXiv:2407.15256.

\bibitem[\protect\citeauthoryear{Lyu, Fan, H{\"u}ser, Hartout, Gumbsch, Faltys,
  Merz, R{\"a}tsch, and Borgwardt}{Lyu et~al.}{2024}]{lyu2024empirical}
Lyu, X., B.~Fan, M.~H{\"u}ser, P.~Hartout, T.~Gumbsch, M.~Faltys, T.~M. Merz,
  G.~R{\"a}tsch, and K.~Borgwardt (2024).
\newblock An empirical study on {KDIGO}-defined acute kidney injury prediction
  in the intensive care unit.
\newblock {\em Bioinformatics\/}~{\em 40\/}(Supplement 1), i247--i256.


\bibitem[\protect\citeauthoryear{Moor, Bennett, Ple{\v{c}}ko, Horn, Rieck,
  Meinshausen, B{\"u}hlmann, and Borgwardt}{Moor
  et~al.}{2023}]{moor2023predicting}
Moor, M., N.~Bennett, D.~Ple{\v{c}}ko, M.~Horn, B.~Rieck, N.~Meinshausen,
  P.~B{\"u}hlmann, and K.~Borgwardt (2023).
\newblock Predicting sepsis using deep learning across international sites: a
  retrospective development and validation study.
\newblock {\em EClinicalMedicine\/}~{\em 62}.


\bibitem[\protect\citeauthoryear{Moukheiber, Temps, Molgi, Li, Lu, Nannapaneni,
  Chahin, Hao, Torres~Fabregas, Celi, Wong, Lloyd, X., Lee, Schneider, Pollard,
  Luo, Kho, and Mark}{Moukheiber et~al.}{2024}]{moukheibert2024northwestern}
Moukheiber, D., W.~Temps, B.~Molgi, Y.~Li, A.~Lu, P.~Nannapaneni, A.~Chahin,
  S.~Hao, F.~Torres~Fabregas, L.~A. Celi, A.~Wong, M.~Lloyd, B.~F. X., H.~Lee,
  D.~Schneider, T.~Pollard, Y.~Luo, A.~Kho, and R.~Mark (2024).
\newblock Northwestern {ICU} ({NWICU}) database (version 0.1.0).
\newblock {\em PhysioNet\/}.


\bibitem[\protect\citeauthoryear{Pollard, Johnson, Raffa, Celi, Mark, and
  Badawi}{Pollard et~al.}{2018}]{pollard2018eicu}
Pollard, T.~J., A.~E. Johnson, J.~D. Raffa, L.~A. Celi, R.~G. Mark, and
  O.~Badawi (2018).
\newblock The {eICU} collaborative research database, a freely available
  multi-center database for critical care research.
\newblock {\em Scientific data\/}~{\em 5\/}(1), 1--13.


\bibitem[\protect\citeauthoryear{Rockenschaub, Hilbert, Kossen, Elbers, von
  Dincklage, Madai, and Frey}{Rockenschaub
  et~al.}{2024}]{rockenschaub2024impact}
Rockenschaub, P., A.~Hilbert, T.~Kossen, P.~Elbers, F.~von Dincklage, V.~Madai,
  and D.~Frey (2024).
\newblock The impact of multi-institution datasets on the generalizability of
  machine learning prediction models in the {ICU}.
\newblock {\em Critical Care Medicine\/}~{\em 52\/}(11), 1710--1721.


\bibitem[\protect\citeauthoryear{Rodemund, Wernly, Jung, Cozowicz, and
  Kok{\"o}fer}{Rodemund et~al.}{2024}]{rodemund2024sicdb}
Rodemund, N., B.~Wernly, C.~Jung, C.~Cozowicz, and A.~Kok{\"o}fer (2024).
\newblock Harnessing big data in critical care: Exploring a new european
  dataset.
\newblock {\em Scientific Data\/}~{\em 11\/}(1), 320.


\bibitem[\protect\citeauthoryear{Roland, B{\"o}ck, Tschoellitsch, Maletzky,
  Hochreiter, Meier, and Klambauer}{Roland et~al.}{2022}]{roland2022domain}
Roland, T., C.~B{\"o}ck, T.~Tschoellitsch, A.~Maletzky, S.~Hochreiter,
  J.~Meier, and G.~Klambauer (2022).
\newblock Domain shifts in machine learning based {Covid-19} diagnosis from
  blood tests.
\newblock {\em Journal of Medical Systems\/}~{\em 46\/}(5), 23.


\bibitem[\protect\citeauthoryear{Rothenh{\"a}usler, Meinshausen, B{\"u}hlmann,
  and Peters}{Rothenh{\"a}usler et~al.}{2021}]{rothenhausler2021anchor}
Rothenh{\"a}usler, D., N.~Meinshausen, P.~B{\"u}hlmann, and J.~Peters (2021).
\newblock Anchor regression: Heterogeneous data meet causality.
\newblock {\em Journal of the Royal Statistical Society Series B: Statistical
  Methodology\/}~{\em 83\/}(2), 215--246.


\bibitem[\protect\citeauthoryear{Sagawa, Koh, Hashimoto, and Liang}{Sagawa
  et~al.}{2020}]{sagawa2019distributionally}
Sagawa, S., P.~W. Koh, T.~B. Hashimoto, and P.~Liang (2020).
\newblock Distributionally robust neural networks.
\newblock In {\em International Conference on Learning Representations}.

\bibitem[\protect\citeauthoryear{Sola, B\"uhlmann, and Shen}{Sola
  et~al.}{2025}]{sola2025causality}
Sola, M., P.~B\"uhlmann, and X.~Shen (2025).
\newblock Causality-inspired robustness for nonlinear models via representation
  learning.
\newblock arXiv:2505.12868.

\bibitem[\protect\citeauthoryear{Thoral, Peppink, Driessen, Sijbrands,
  Kompanje, Kaplan, Bailey, Kesecioglu, Cecconi, Churpek, et~al.}{Thoral
  et~al.}{2021}]{thoral2021sharing}
Thoral, P.~J., J.~M. Peppink, R.~H. Driessen, E.~J. Sijbrands, E.~J. Kompanje,
  L.~Kaplan, H.~Bailey, J.~Kesecioglu, M.~Cecconi, M.~Churpek, et~al. (2021).
\newblock Sharing {ICU} patient data responsibly under the society of critical
  care medicine/{European} society of intensive care medicine joint data
  science collaboration: the {Amsterdam} university medical centers database
  ({AmsterdamUMCdb}) example.
\newblock {\em Critical care medicine\/}~{\em 49\/}(6), e563--e577.


\bibitem[\protect\citeauthoryear{Toma{\v{s}}ev, Glorot, Rae, Zielinski, Askham,
  Saraiva, Mottram, Meyer, Ravuri, Protsyuk, et~al.}{Toma{\v{s}}ev
  et~al.}{2019}]{tomavsev2019clinically}
Toma{\v{s}}ev, N., X.~Glorot, J.~W. Rae, M.~Zielinski, H.~Askham, A.~Saraiva,
  A.~Mottram, C.~Meyer, S.~Ravuri, I.~Protsyuk, et~al. (2019).
\newblock A clinically applicable approach to continuous prediction of future
  acute kidney injury.
\newblock {\em Nature\/}~{\em 572\/}(7767), 116--119.


\bibitem[\protect\citeauthoryear{Ulmer and Scheidegger}{Ulmer and
  Scheidegger}{2025}]{ulmer2025anchorforest}
Ulmer, M. and C.~Scheidegger (2025).
\newblock \texttt{AnchorForest} of \texttt{SDModels}.
\newblock Vignette, R package.
\newblock Available at
  \href{https://www.markus-ulmer.ch/SDModels/articles/AnchorForest.html}{\texttt{www.markus-ulmer.ch/SDModels/articles/AnchorForest.html}}.

\bibitem[\protect\citeauthoryear{Ulmer, Scheidegger, and Bühlmann}{Ulmer
  et~al.}{2025}]{ulmer2025spectrally}
Ulmer, M., C.~Scheidegger, and P.~Bühlmann (2025).
\newblock Spectrally deconfounded random forests.
\newblock {\em Journal of Computational and Graphical Statistics\/}, 1--11.


\bibitem[\protect\citeauthoryear{van~de Water, Schmidt, Elbers, Thoral,
  Arnrich, and Rockenschaub}{van~de Water et~al.}{2024}]{water2023yet}
van~de Water, R., H.~N.~A. Schmidt, P.~Elbers, P.~Thoral, B.~Arnrich, and
  P.~Rockenschaub (2024).
\newblock Yet another {ICU} benchmark: A flexible multi-center framework for
  clinical {ML}.
\newblock In {\em The Twelfth International Conference on Learning
  Representations}.

\bibitem[\protect\citeauthoryear{Xu, Chen, Zhu, Yu, Chen, Huang, Wu, and
  Zhang}{Xu et~al.}{2022}]{xu2022critical}
Xu, P., L.~Chen, Y.~Zhu, S.~Yu, R.~Chen, W.~Huang, F.~Wu, and Z.~Zhang (2022).
\newblock Critical care database comprising patients with infection.
\newblock {\em Frontiers in Public Health\/}~{\em 10}, 852410.


\bibitem[\protect\citeauthoryear{Yang, Soltan, and Clifton}{Yang
  et~al.}{2022}]{yang2022machine}
Yang, J., A.~A. Soltan, and D.~A. Clifton (2022).
\newblock Machine learning generalizability across healthcare settings:
  insights from multi-site {COVID-19} screening.
\newblock {\em npj Digital Medicine\/}~{\em 5\/}(1), 69.


\bibitem[\protect\citeauthoryear{Y{\`e}che, Kuznetsova, Zimmermann, H{\"u}ser,
  Lyu, Faltys, and R\"atsch}{Y{\`e}che et~al.}{2021}]{yeche2021hirid}
Y{\`e}che, H., R.~Kuznetsova, M.~Zimmermann, M.~H{\"u}ser, X.~Lyu, M.~Faltys,
  and G.~R\"atsch (2021).
\newblock Hi{RID}-{ICU}-benchmark --- a comprehensive machine learning
  benchmark on high-resolution {ICU} data.
\newblock In {\em Thirty-fifth Conference on Neural Information Processing
  Systems Datasets and Benchmarks Track}.

\end{thebibliography}

\appendix

\section{Details on the data}
\label{app:data}

\begin{landscape}
\newgeometry{left=3cm, right=1cm, top=10cm, bottom=0cm}
\begin{table*}[htb!]
\centering
\caption{
\label{tab:dataset_summaries}
Dataset summaries and outcome statistics.}
\vspace{0.2cm}
\small
\begin{tabular}{lrrrrrrrrr}
 & AUMCdb & eICU & HIRID & MIMIC-III (CV) & MIMIC-IV & SICdb & NWICU & PICdb & Zigong \\
\midrule
country & Netherlands & USA & Switzerland & USA & USA & Austria & USA & China & China \\
years & 2003--2016 & 2015--2016 & 2008--2016 & 2001--2008 & 2008--2022 & 2013--2021 & 2020--2022 & 2010--2018 & 2019--2020 \\
\midrule
num. patients & 19,993 & 188,257 & 33,586 & 27,337 & 65,204 & 21,403 & 22,969 & 12,565 & 2,583 \\
num. stays & 22,897 & 188,257 & 33,586 & 34,154 & 93,679 & 27,115 & 28,150 & 13,516 & 2,583 \\
average LoS & 3.3 days & 2.7 days & 2.2 days & 4.3 days & 3.3 days & 3.0 days & 3.2 days & 5.2 days & 6.5 days \\
\midrule
\multicolumn{4}{l}{\emph{circulatory failure within 8 hours}}  \\
num. patients & 8,129 & 29,671 & 24,809 & 8,032 & 30,149 & 19,802 & 3,138 & 10,341 & 1,872 \\
num. samples & 217,280 & 288,092 & 607,186 & 145,401 & 565,139 & 1,125,649 & 48,829 & 479,691 & 44,447 \\
prevalence & 9.3\% & 17.8\% & 5.2\% & 13.4\% & 13.2\% & 3.9\% & 15.1\% & 1.2\% & 4.6\% \\
\midrule
\multicolumn{4}{l}{\emph{acute kidney injury within 48 hours}}\\
num. patients & 17,037 & 115,994 & 13,572 & 18,996 & 54,670 & 17,534 & 17,319 & 5,472 & 2,125 \\
num. samples & 715,574 & 3,993,291 & 504,071 & 1,057,755 & 2,616,961 & 709,211 & 800,926 & 205,776 & 101,236 \\
prevalence & 3.9\% & 10.2\% & 5.2\% & 5.0\% & 8.1\% & 4.1\% & 4.3\% & 3.8\% & 3.6\% \\
\midrule
\multicolumn{4}{l}{\emph{log(lactate in 4 hours)}} \\
num. patients & 9,169 & 36,132 & 28,127 & 9,451 & 33,915 & 20,687 & 5,017 & 10,794 & 2,143 \\
num. samples & 103,939 & 97,535 & 188,768 & 51,492 & 236,291 & 466,219 & 43,142 & 108,337 & 10,025 \\
mean (sd) & 0.48 (0.62) & 0.74 (0.75) & 0.44 (0.58) & 0.76 (0.69) & 0.72 (0.65) & 0.32 (0.47) & 0.66 (0.69) & 0.46 (0.63) & 0.74 (0.61) \\
\midrule
\multicolumn{4}{l}{\emph{log(creatinine in 24 hours)}} \\
num. patients & 8,946 & 97,458 & 11,719 & 16,277 & 43,399 & 12,085 & 14,426 & 5,053 & 1,826 \\
num. samples & 70,312 & 377,230 & 39,636 & 110,551 & 335,222 & 53,039 & 99,836 & 11,826 & 6,222 \\
mean (sd) & 0.22 (0.59) & 0.23 (0.66) & 0.12 (0.58) & 0.23 (0.68) & 0.27 (0.65) & 0.12 (0.53) & 0.25 (0.64) & -0.38 (0.56) & 0.05 (0.64) \\
\bottomrule
\end{tabular}
\end{table*}
\restoregeometry
\end{landscape}

\subsection{Tasks}
\label{app:data:tasks}
We give additional details to \cref{sec:data:outcomes}.

\paragraph{Details on circulatory failure}
Recall that a patient is experiencing circulatory failure if they have low mean arterial blood pressure (the patient's mean arterial pressure is below 65 mmHg or the patient is receiving treatment to elevate blood pressure) and high lactate (above 2 mmol/l).
We assign a positive event only if both the (i) high lactate and (ii) the low blood pressure conditions are satisfied.
Similarly, we only assign a negative event if both (i) and (ii) are negative.
That is, for a certain time, a patient has a measured value of lactate below 2 mmol/l and a measured value of blood pressure above 65 mmHg.
Blood pressure is a vital sign, so it has a very low missingness rate for the core datasets, and the above definition results in reasonable event labels.
However, for PICdb and Zigong, mean arterial blood pressure can be extracted for only very few time points, leading to almost only positive events where patients have high lactate and receive medication to suppress blood pressure.
Thus, for these targets, we define negative events as (i) high lactate and (ii) the patient does not receive medication to suppress blood pressure and blood pressure is above 65mmHg or not measured.

\paragraph{Details on acute kidney injury}
Acute kidney injury is defined as AKI 3 according to the KDIGO guidelines \citep{kidney2012kdigo}.
These are: (i) A patient has an increase in creatinine of a factor of 3 relative to their 7 day baseline; (ii) A patient has acute kidney injury level 1 according to the KDIGO guidelines and has a creatinine value of at least 4.0 mmol/L; (iii) A patient has anuria, that is, has not produced urine over the last 12 hours; (iv) A patient had an average relative urine rate below 0.3ml/kg/h over the last 24 hours; (v) The patient has started renal replacement therapy.
We say that a patient is experiencing an acute kidney injury event if at least one of the conditions (i) - (v) apply.
As urine rate is only measured sparsely for NWICU, PICdb, and Zigong, we define a negative kidney injury event as (i, ii) creatinine is measured and low, (iii, iv) urine rate is measured and normal, or not measured, (v) the patient is not receiving renal replacement therapy.
Again, requiring that urine rate is measured and normal would result in very few negative events for the datasets where urine rate is only very sparsely measured.

\subsection{Feature engineering}
\label{app:data:feature_engineering}
Both linear and tree-based methods do not natively handle time-series data.
We therefore compute features to summarize the patient's history.
Many variables in the ICU are long-tailed. We log-transform these before the feature engineering.

For continuous variables, we compute (i) the last observed value filled forwards, (ii) the square of the last observed value filled forwards, and (iii) an indicator of whether the feature was missing. Then for a task-specific horizon, we compute (iv) the mean, (v) the standard deviation, (vi) the minimum, (vii) the maximum, (viii) the slope of a linear fit, (ix) the fraction of nonmissing values, and (x) an indicator whether all values in the horizon were missing.
For discrete variables, we compute: (i) the last value and, for a task-specific horizon, the (ii) mode and (iii) the fraction of non-missing values.
For treatment indicators, we compute (i) the last value and, for a task-specific horizon, (ii) the fraction of positive values over the past horizon and (iii) an indicator whether there were any positive values in the past horizon.
For continuous treatment variables, for a task-specific horizon, we compute (i) the logarithm of the rate, that is, the logarithm of the average amount administered per minute.
We include the treatment indicator whenever we include the rate.

We use an 8 hour horizon for feature engineering for the prediction tasks of log(lactate) and circulatory failure and a 24 hour horizon for the feature engineering for the prediction tasks of log(creatinine), acute kidney injury.

The feature engineering is defined in \href{https://github.com/eth-mds/icu-features}{\texttt{github.com/eth-mds/icu-features}}.

\subsection{Subselection of variables}
\label{app:data:variable_selection}
For prediction of log(lactate) and circulatory failure, we use the top 20 variables of \citet{hyland2020early} according to their table 1.
These are: age, time in hours since ICU admission,
the Richmond agitation sedation score,
heart rate, diastolic blood pressure, mean arterial pressure, systolic blood pressure, pulse oximeter oxygen saturation (SpO2),
cardiac output, c-reactive protein, serum glucose, lactate,  normalized prothrombin time,
an indicator whether a patient is receiving any circulatory failure treatment, the treatment rate and an indicator for dobutamine, levosimendan, milrinone, and theophylline, the peak pressure of mechanical ventilation, an indicator of non-opioid pain medication, and supplemental oxygen from ventilation.

For the prediction of log(creatinine) and acute kidney injury, we use the top variables of \citet{lyu2024empirical} according to their figure 8a.
These are:
weight, time in hours since ICU admission,
creatinine, end-tidal CO2, c-reactive protein, respiratory rate, bilirubin, magnesium, potassium,
relative urine rate [ml/kg/h], rates and indicators for ultrafiltration on continuous RRT, heparin, and loop diuretics, indicator for fluid administration, anicoagulant treatments, antidelirium treatment, opioid pain medication, antibiotics, and ventilation.

\section{Details on anchor boosting}

\subsection{Details on the probit anchor loss}
\label{app:probit_anchor_loss}
Recall that $\Phi$ and $\phi$ are the Gaussian distribution's cumulative distribution function and probability density function.
For scores $f \in \mathbb{R}^n$ and outcome $y \in \{-1, 1\}^n$, the negative log-likelihood is $- \sum_{i=1}^n \log( \Phi(y_i f_i))$ with gradient $r(f) := -y \cdot \phi(f) / \Phi(yf)$.
We use the gradient as (score) residuals \citep{kook2022distributional}.
The probit anchor loss with parameter $\gamma$ is
\begin{equation*}
    \ell(f, y) = -\sum_{i=1}^n \log( \Phi(y_i f_i) ) + \frac{1}{2} (\gamma - 1) \| P_A r(f) \|^2
\end{equation*}
Calculate $\dot{r}(f) := \frac{\dd}{\dd f} r(f) = -f \phi(f) r + r^2$ and $\ddot{r}(f) := \frac{\dd^2}{\dd f^2} r(f) = (f^2 - 1)  r - 3 f r^2 + 2 r^3$. Then,
\begin{equation}
    \label{eq:probit_loss_grad}
    g(f, y) = r + (\gamma - 1) P_A r \cdot \dot{r}(f)
\end{equation}
and
\begin{equation}
    \label{eq:probit_loss_hess}
    H(f, y) = \diag( \dot{r}(f) + (\gamma - 1) P_A r \cdot \ddot{r}(f) ) + (\gamma - 1) \cdot \diag( \dot{r}(f) ) P_A \diag( \dot{r}(f) ).
\end{equation}

\subsection{The logistic anchor loss is not convex}
\label{app:logistic_anchor_is_non_convex}
\begin{figure}[H]
    \centering
    \includegraphics[width=0.8\textwidth]{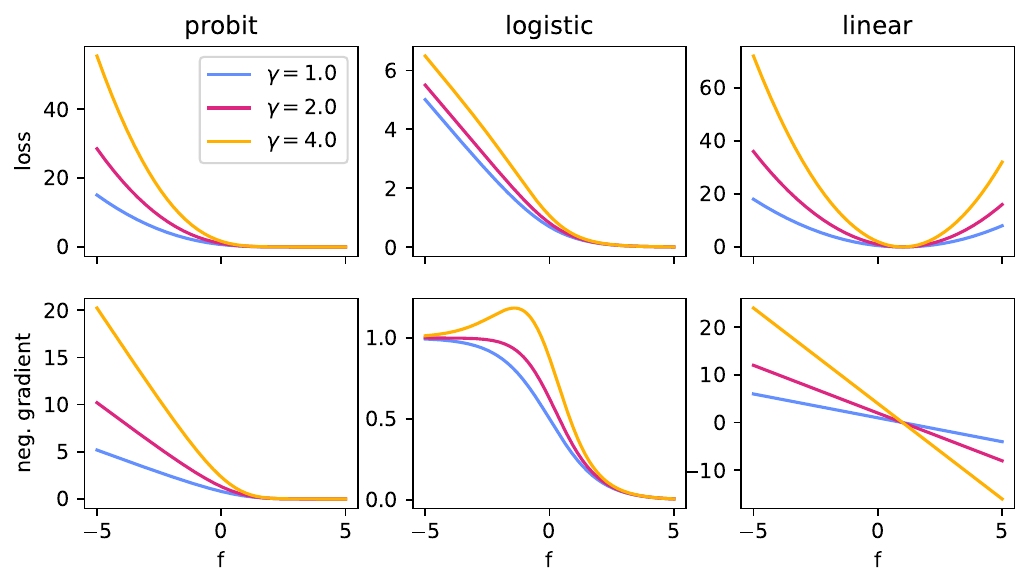}
    \caption{
        \label{fig:anchor_losses}
        Different anchor losses for a single observation with $y = 1$.
    }
\end{figure}

Write $\sigma(f) = 1/ (1 + e^{-f})$.
For scores $f \in \mathbb{R}^n$ and outcomes $y \in \{-1, 1\}^n$, the negative log-likelihood using the logistic link $\sigma$ is $- \sum_{i=1}^n \log( \sigma(y_i f_i))$ with gradient $r(f) := y \cdot \sigma(-yf)$.
Using the gradient as (score) residuals \citep{kook2022distributional}, the logistic anchor loss with parameter $\gamma$ is
\begin{equation*}
    \ell(f, y) = -\sum_{i=1}^n \log( \sigma(y_i f_i) ) + \frac{1}{2} (\gamma - 1) \| P_A r(f) \|^2
\end{equation*}
Calculate $\dot{r}(f) := \frac{\dd}{\dd f} r(f) = \sigma(f) \cdot \sigma(-f)$ and $\ddot{r}(f) := \frac{\dd^2}{\dd^2 f} r(f) = y \cdot \sigma(f) \sigma(-f) \cdot (1 - 2 \sigma(-f))$.
\Cref{eq:probit_loss_grad,eq:probit_loss_hess} also apply.
In \cref{fig:anchor_losses}, we show different anchor losses for $\gamma = 1, 2, 4$.
For $\gamma = 4$, the logistic anchor loss is visibly non-convex.

\subsection{Second order tree node value optimization matters}
\label{app:second_order_tree_node_value_optimization}
Initially, we implemented anchor boosting without the second-order optimization of the anchor loss for the tree node values.
That is, we used the gradient's mean as the tree's leaf values (as if the Hessian was the identity) or the gradient's mean divided by the sum of the Hessian that would result from the loss with $\gamma = 1$.
These are the same for regression.

The algorithm still converged, but, as the anchor loss scales with $\gamma$, we needed to use a much smaller learning rate.
The optimal learning rate depended strongly on $\gamma$ and needed to be tuned.
This required more trees, and the tuning resulted in higher variance.
Finally, even for very small learning rates $\mathrm{lr} \leq 0.001$, the algorithm diverged whenever $\gamma \geqslant 50$.
The second-order optimization of the tree node values we implemented solves these problems.
The loss of efficiency is made up by the smaller number of trees required.
The resulting algorithm is also much more robust to the learning rate, which, as is the case for standard tree-based gradient-boosting, does not need to be tuned.

\FloatBarrier
\section{Additional figures}
\FloatBarrier

\label{app:additional_figures}
\setlength{\abovecaptionskip}{1pt} 
\setlength{\belowcaptionskip}{1pt} 
\setlength{\textfloatsep}{1pt }
\begin{figure}[ht]
    \includegraphics[width=\textwidth]{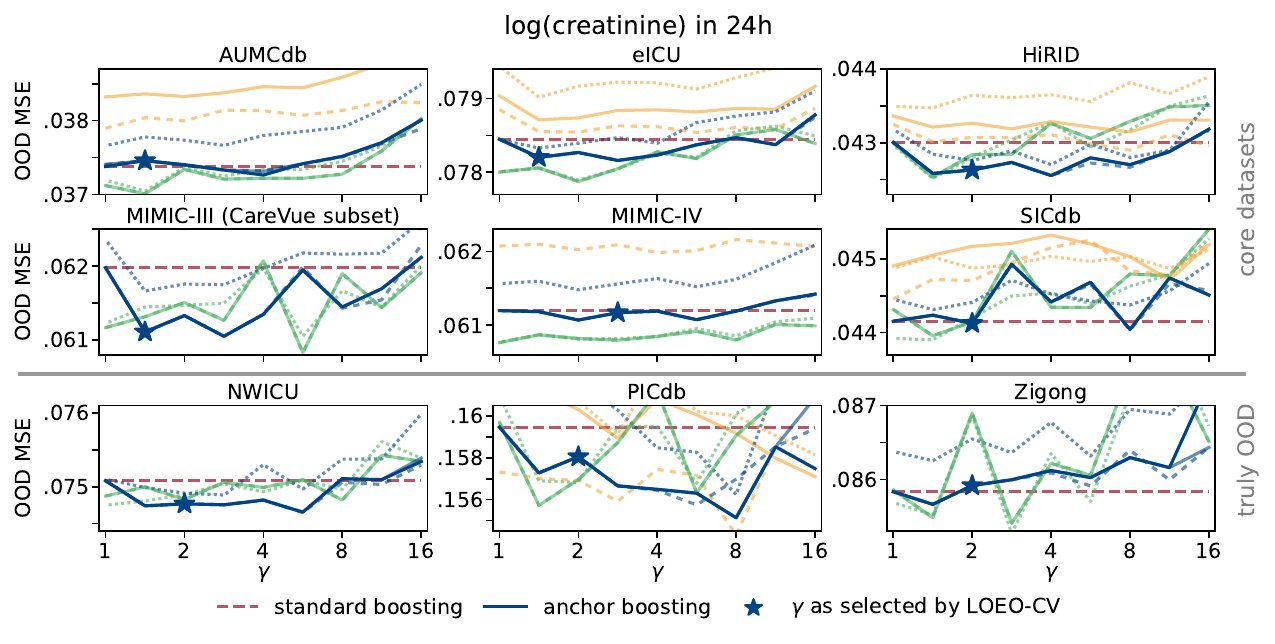}
    \caption{
        \label{fig:crea_algbm_tune}
        Boosted anchor regression's OOD MSE predicting log(creatinine) in 24 hours as a function of $\gamma$.
        We vary the number of trees from 500 (dotted), 1000 (solid), to 2000 (dashed) and the trees' maximal depth from 2 (orange), 3 (blue), to 4 (green).
    }
\end{figure}

\begin{figure}[ht]
    \includegraphics[width=\textwidth]{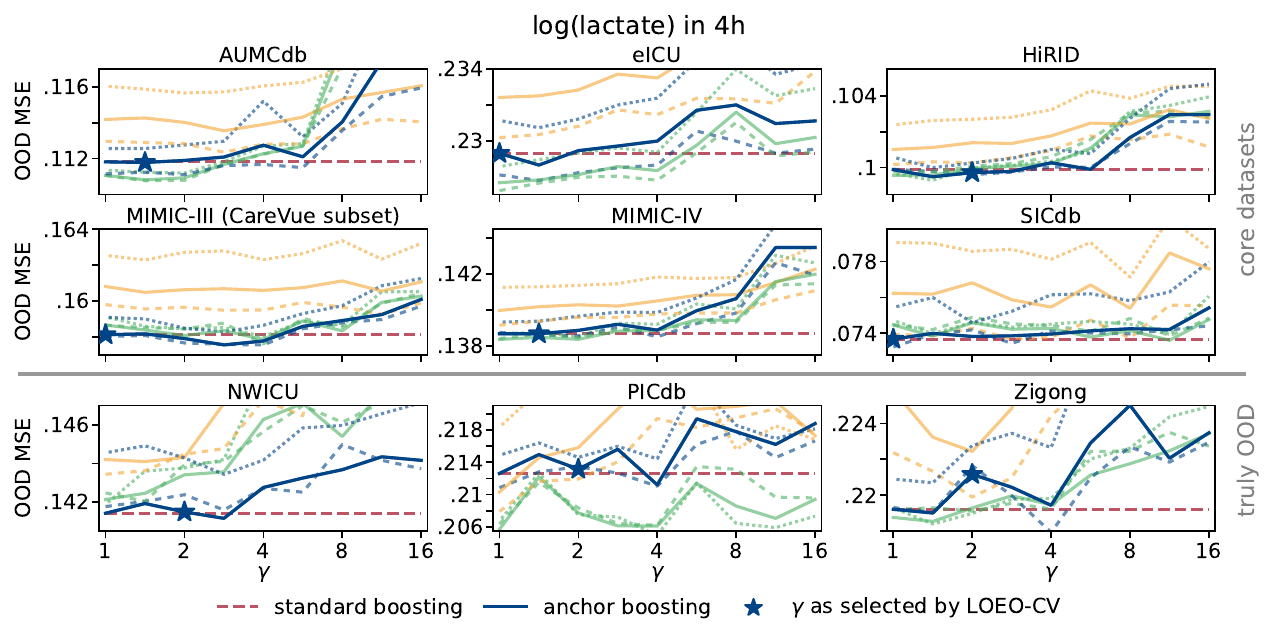}
    \caption{
        \label{fig:lact_algbm_tune}
        Boosted anchor classification's OOD MSE predicting log(lactate) in 4 hours as a function of $\gamma$.
        We vary the number of trees from 500 (dotted), 1000 (solid), to 2000 (dashed) and the trees' maximal depth from 2 (orange), 3 (blue), to 4 (green).
    }
\end{figure}

\begin{figure}[tbph]
    \includegraphics[width=\textwidth]{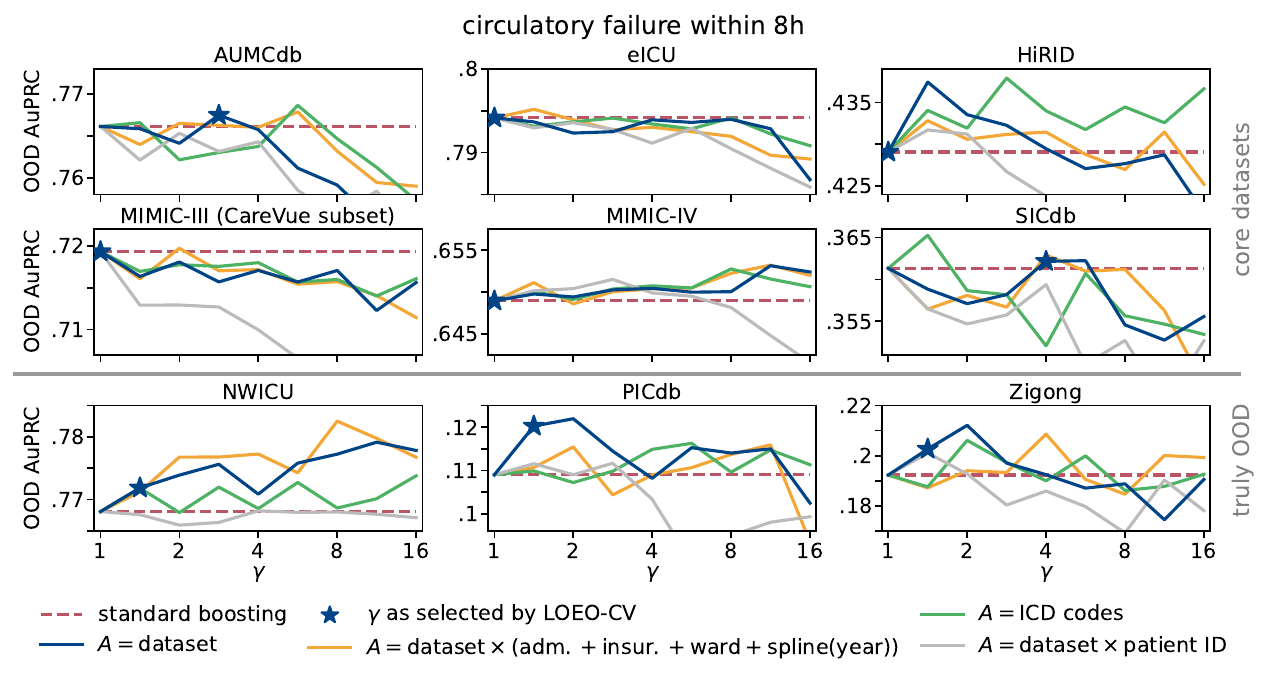}    \caption{
        \label{fig:circ_algbm_which_anchor}
        Boosted anchor classification's OOD AuPRC (larger is better) predicting circulatory failure within 8 hours as a function of $\gamma$ and the anchor used.
    }
\end{figure}

\begin{figure}[tbph]
    \includegraphics[width=\textwidth]{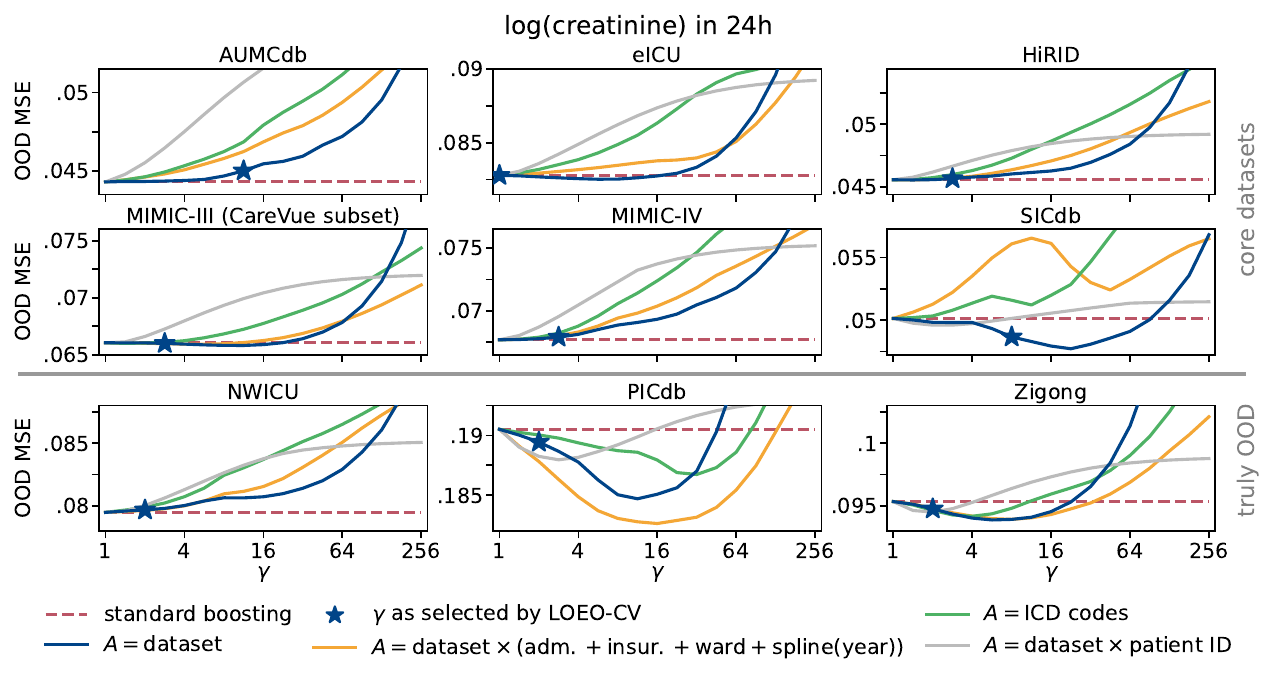}    \caption{
        \label{fig:crea_anchor_which_anchor}
        Linear anchor regression's OOD MSE predicting log(creatinine) in 24 hours as a function of $\gamma$ and the anchor used.
    }
\end{figure}

\begin{figure}[tbph]
    \includegraphics[width=\textwidth]{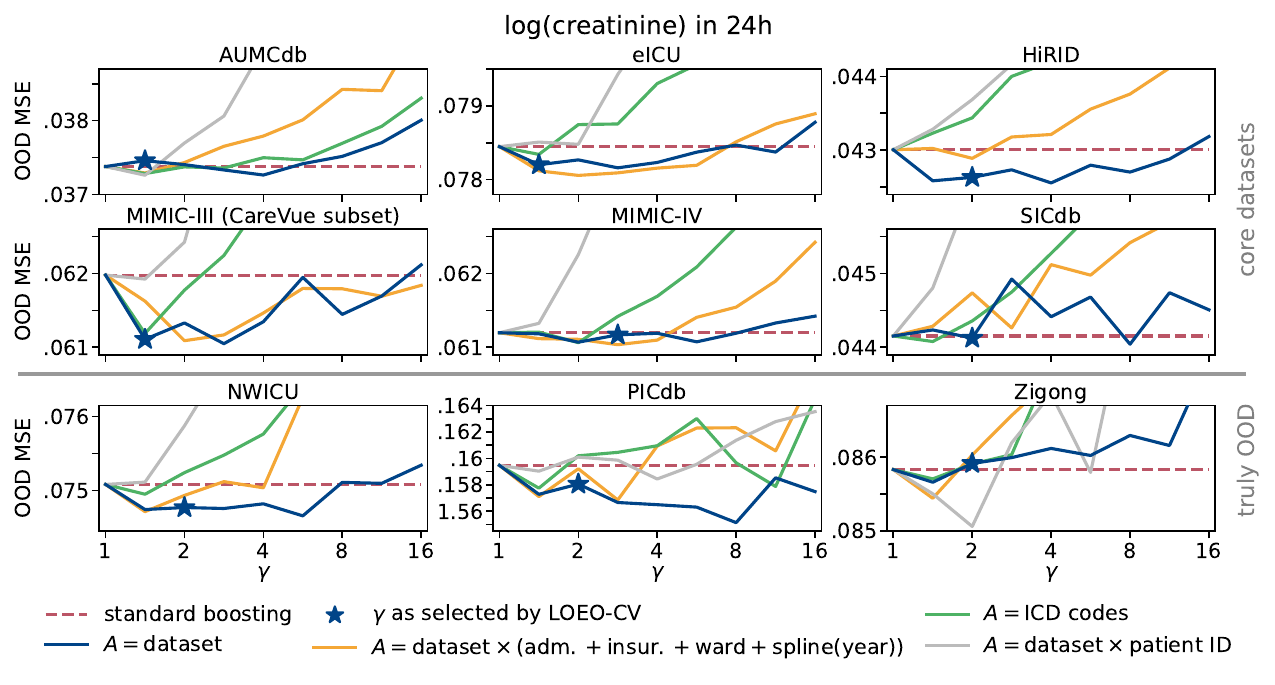}    \caption{
        \label{fig:crea_algbm_which_anchor}
        Boosted anchor regression's OOD MSE predicting log(creatinine) in 24 hours as a function of $\gamma$ and the anchor used.
    }
\end{figure}

\begin{figure}[tbph]
    \includegraphics[width=\textwidth]{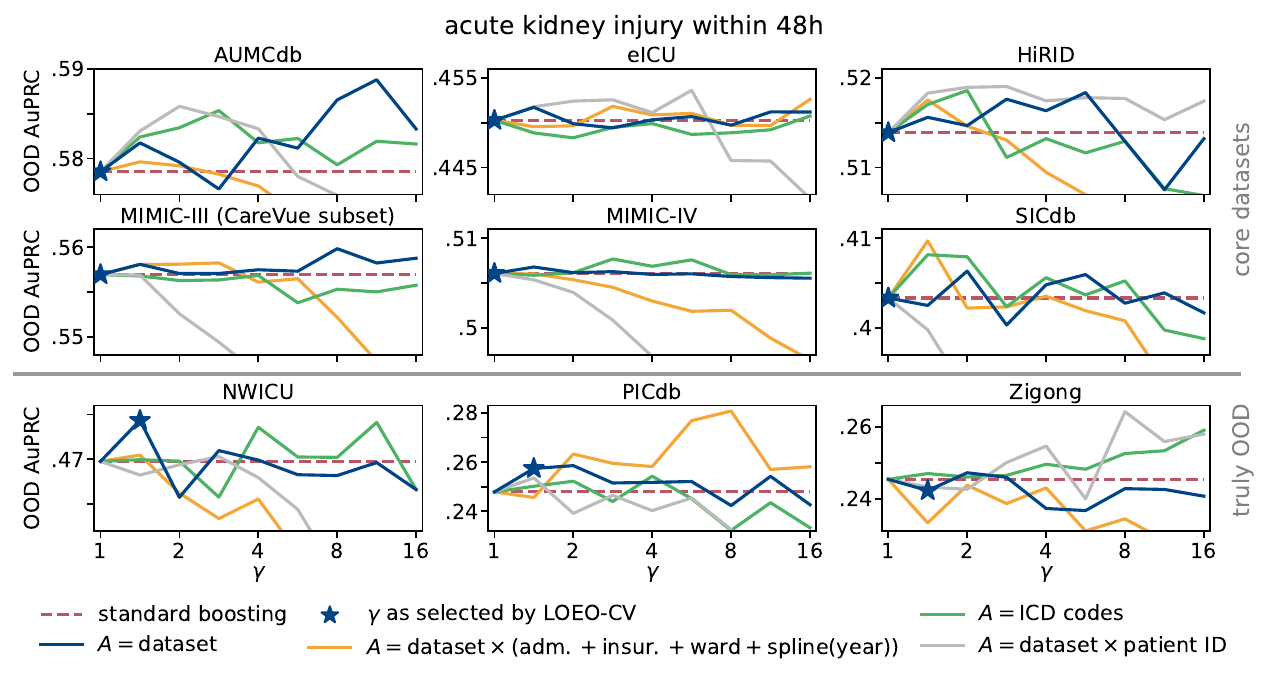}   
    \caption{
        \label{fig:kidney_algbm_which_anchor}
        Boosted anchor classification's OOD AuPRC (larger is better) predicting acute kidney injury within 48 hours as a function of $\gamma$ and the anchor used.
    }
\end{figure}

\begin{figure}[tbph]
    \includegraphics[width=\textwidth]{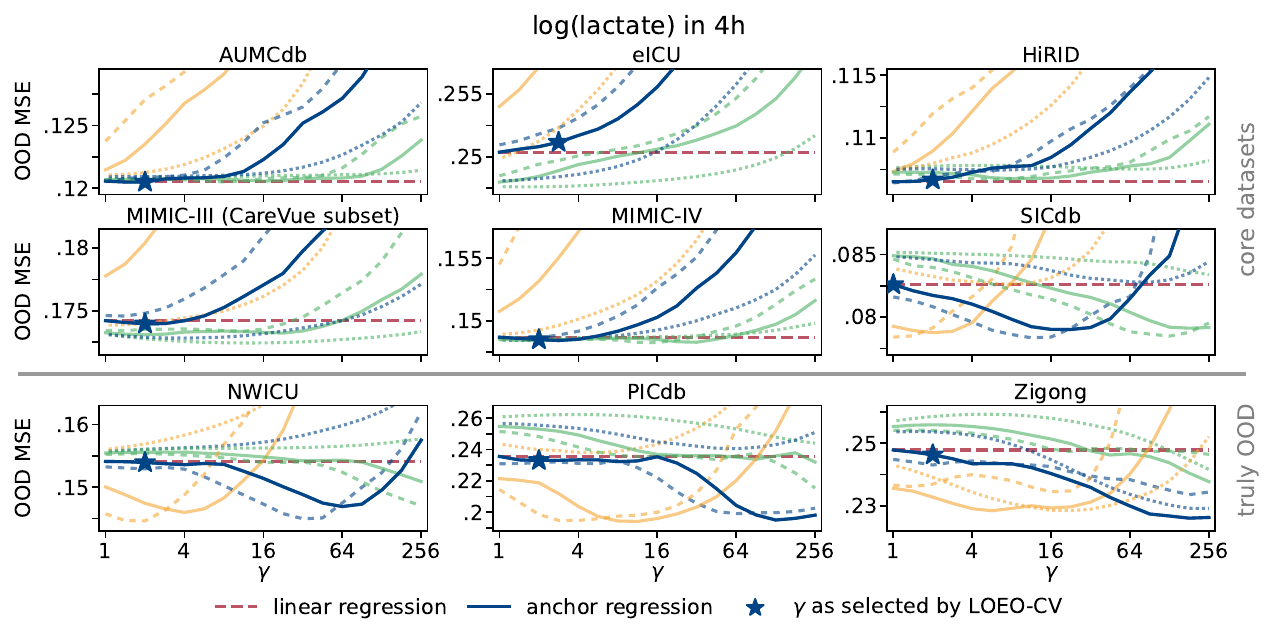}
    \caption{
        \label{fig:lact_anchor_colored}
        Linear anchor regression's OOD MSE predicting log(lactate) in 4 hours as a function of $\gamma$.
        We add an elastic-net regularization term $\lambda \left( \eta \|\beta\|_1 + (1 - \eta) \| \beta \|_2^2 \right)$ to \cref{eq:anchor,eq:anchor_general}.
        Performances are colored by $\lambda = \lambda_\mathrm{max} / 10^2$ (orange), $\lambda_\mathrm{max} / 10^3$ (blue), and $\lambda_\mathrm{max} / 10^4$ (green). %
        Lasso ($\eta=1$) is dashed, elastic net ($\eta = 0.5)$ solid, and ridge ($\eta=0$) dotted.
    }
\end{figure}

\begin{figure}[tbph]
    \centering
    \hspace{1cm}
    \includegraphics[width=0.85\textwidth]{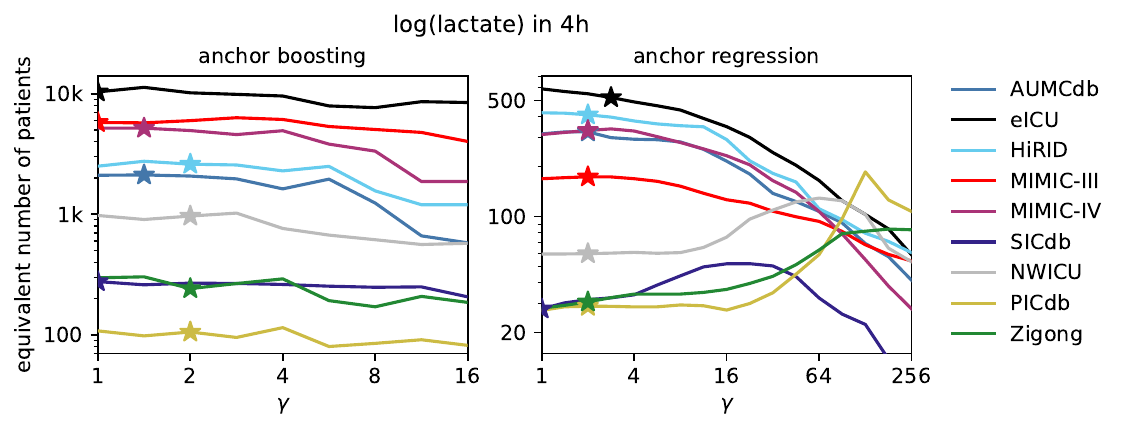}
    \caption{
        \label{fig:lact_rescaled}
        OOD performances predicting log(lactate) in 4 hours as a function of $\gamma$, rescaled by the number of patients from the target domain required to match that performance.
    }
\end{figure}

\begin{figure}[tbph]
    \centering
    \hspace{1cm}
    \includegraphics[width=0.85\textwidth]{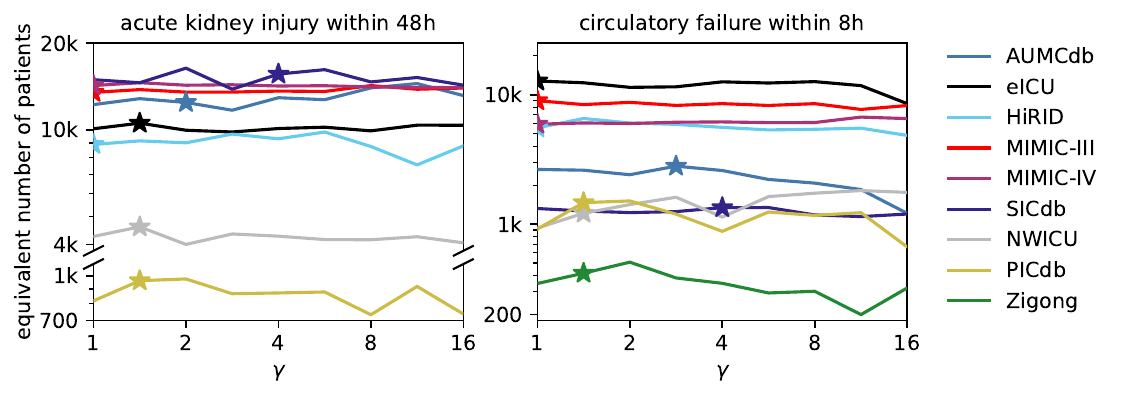}
    \caption{
        \label{fig:classification_rescaled}
        Anchor boosting's OOD performances predicting acute kidney injury in 48 hours and circulatory failure in 8 hours as a function of $\gamma$, rescaled by the number of patients from the target domain required to match that performance.
    }
\end{figure}

\begin{figure}[tbph]
    \includegraphics[width=\textwidth]{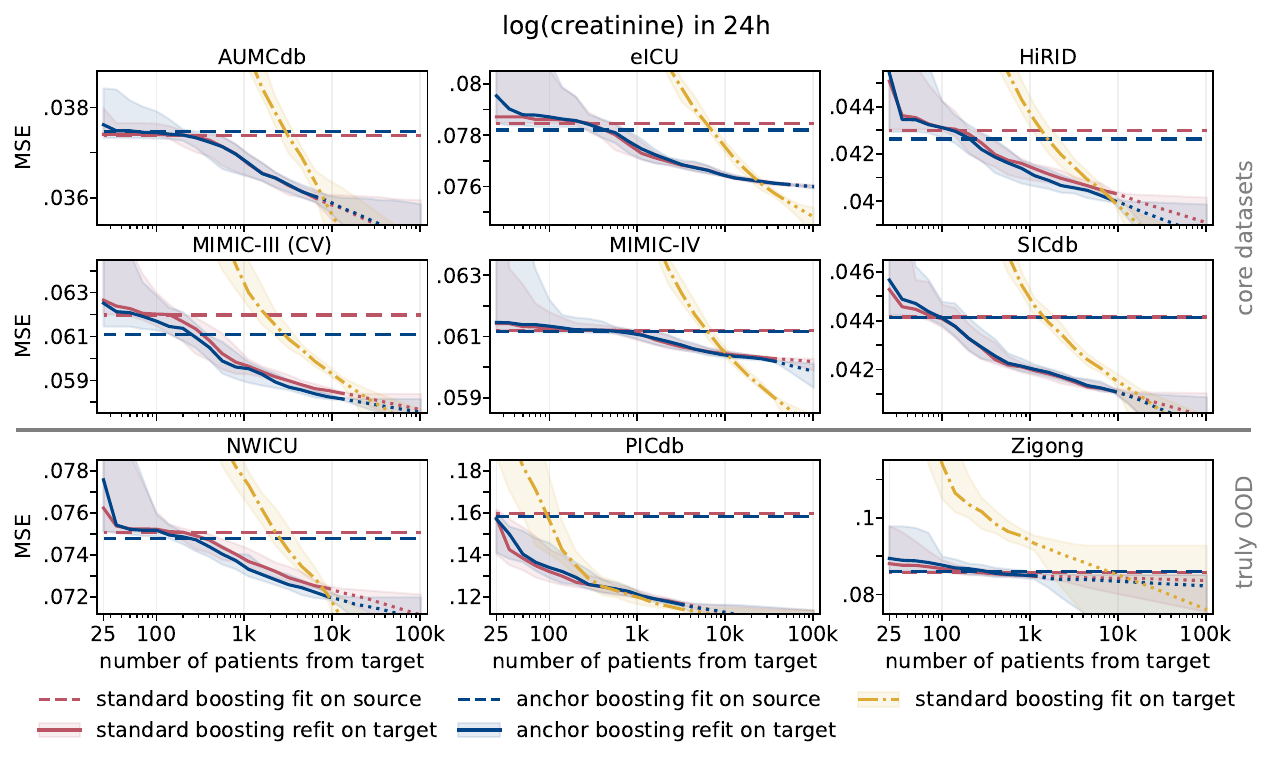}
    \caption{
        \label{fig:crea_algbm_n_samples}
        MSE predicting log(creatinine) in 24 hours as a function of available patients from the target dataset.
        Lines are medians and shaded  areas are 80\% credible sets over 20 different subsampling seeds.
    }
\end{figure}

\begin{figure}[tbph]
    \includegraphics[width=\textwidth]{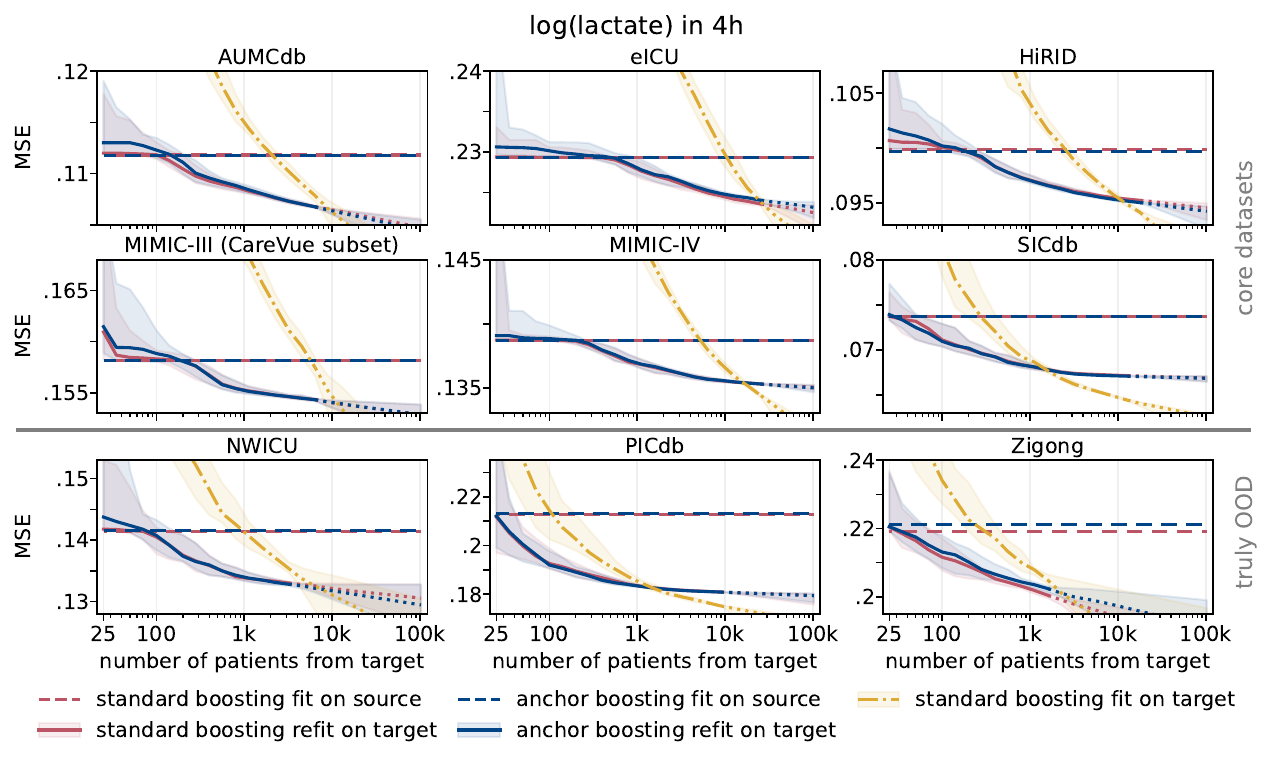}
    \caption{
        \label{fig:lact_algbm_n_samples}
        MSE predicting log(lactate) in 4 hours as a function of available patients from the target dataset.
        Lines are medians and shaded areas are 80\% credible sets over 20 different subsampling seeds.
    }
\end{figure}

\begin{figure}[tbph]
    \includegraphics[width=\textwidth]{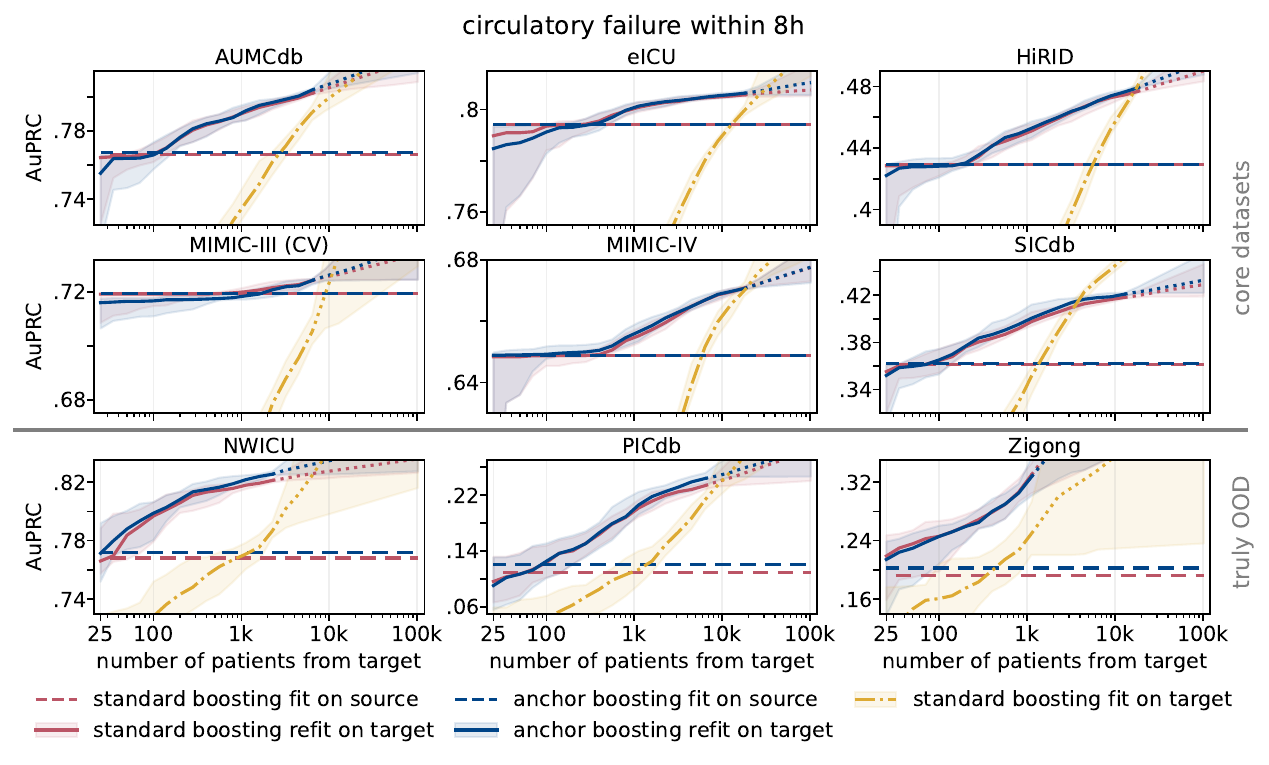}
    \caption{
        \label{fig:circ_algbm_n_samples}
        AuPRC (larger is better) predicting circulatory failure within 8 hours as a function of available patients from the target dataset.
        Lines are medians and shaded areas are 80\% credible sets over 20 different subsampling seeds.
    }
\end{figure}

\begin{figure}[tbph]
    \includegraphics[width=\textwidth]{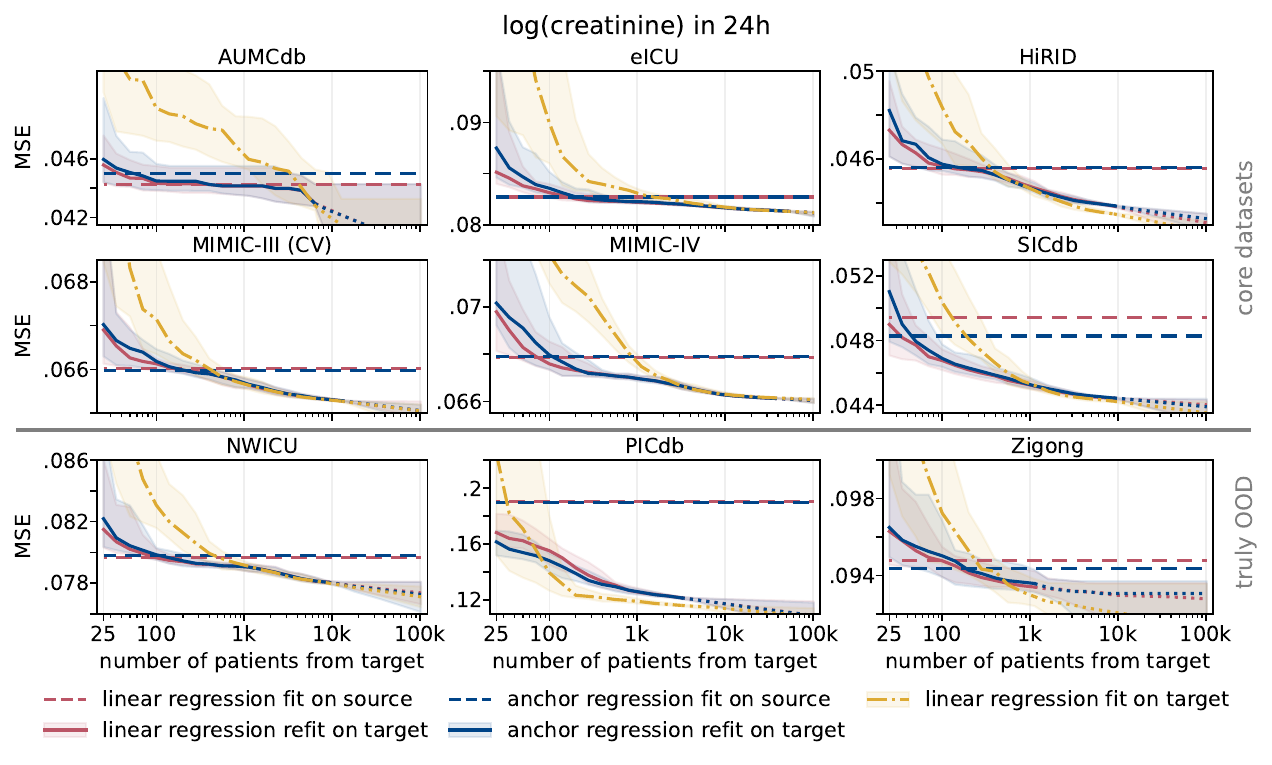}
    \caption{
        \label{fig:crea_anchor_n_samples}
        MSE predicting log(creatinine) in 24 hours as a function of available patients from the target dataset.
        Lines are medians and shaded areas are 80\% credible sets over 20 different subsampling seeds.
    }
\end{figure}

\begin{figure}[tbph]
    \includegraphics[width=\textwidth]{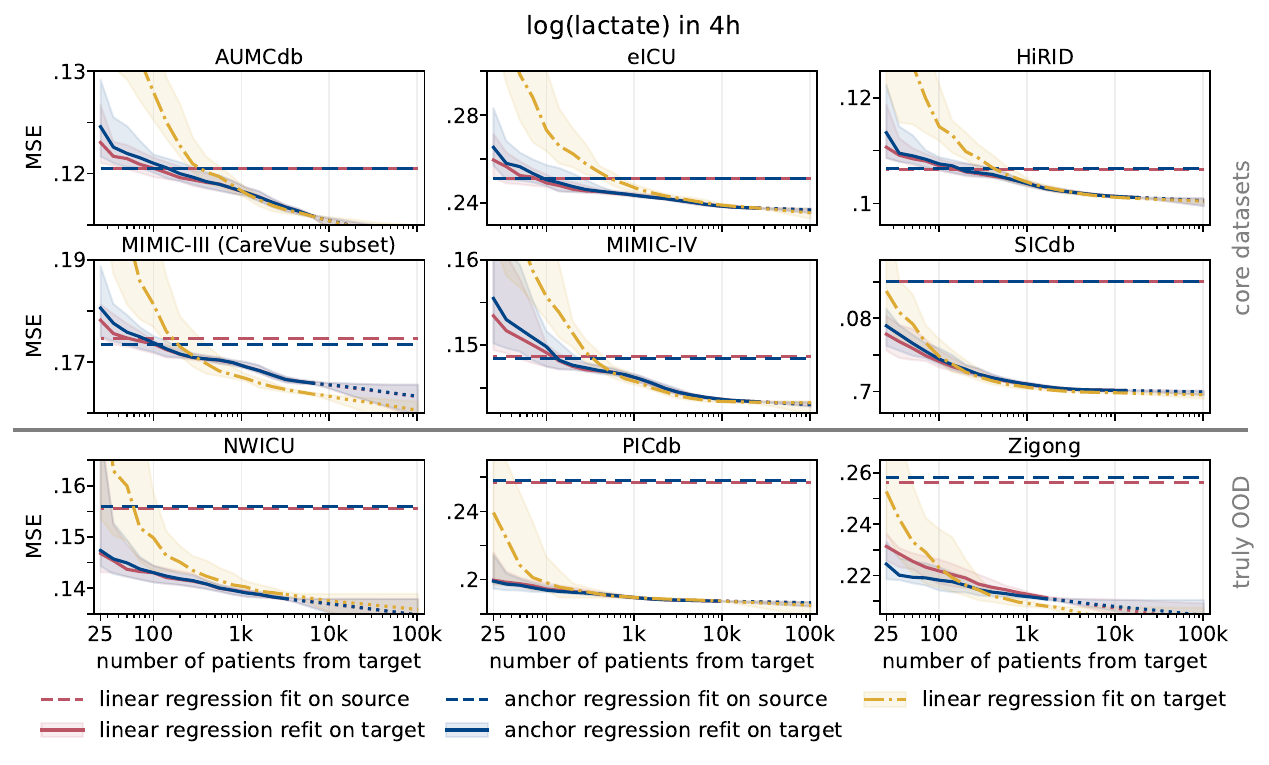}
    \caption{
        \label{fig:lact_anchor_n_samples}
        MSE predicting log(lactate) in 4 hours as a function of available patients from the target dataset.
        Lines are medians and shaded areas are 80\% credible sets over 20 different subsampling seeds.
    }
\end{figure}

\begin{figure}[tbph]
    \includegraphics[width=\textwidth]{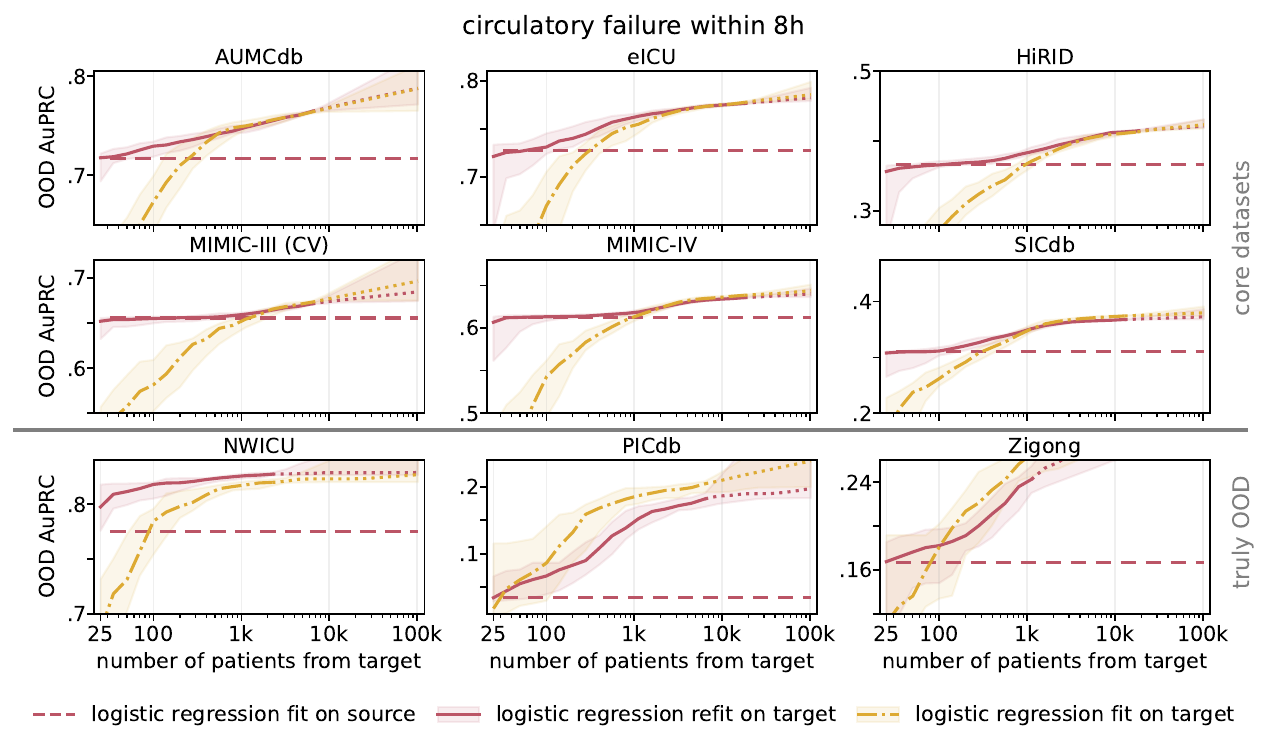}
    \caption{
        \label{fig:circ_glm_n_samples}
        AuPRC (larger is better) predicting circulatory failure within 8 hours as a function of available patients from the target dataset using logistic regression.
        Lines are medians and shaded areas are 80\% credible sets over 20 different subsampling seeds.
    }
\end{figure}

\begin{figure}[tbph]
    \includegraphics[width=\textwidth]{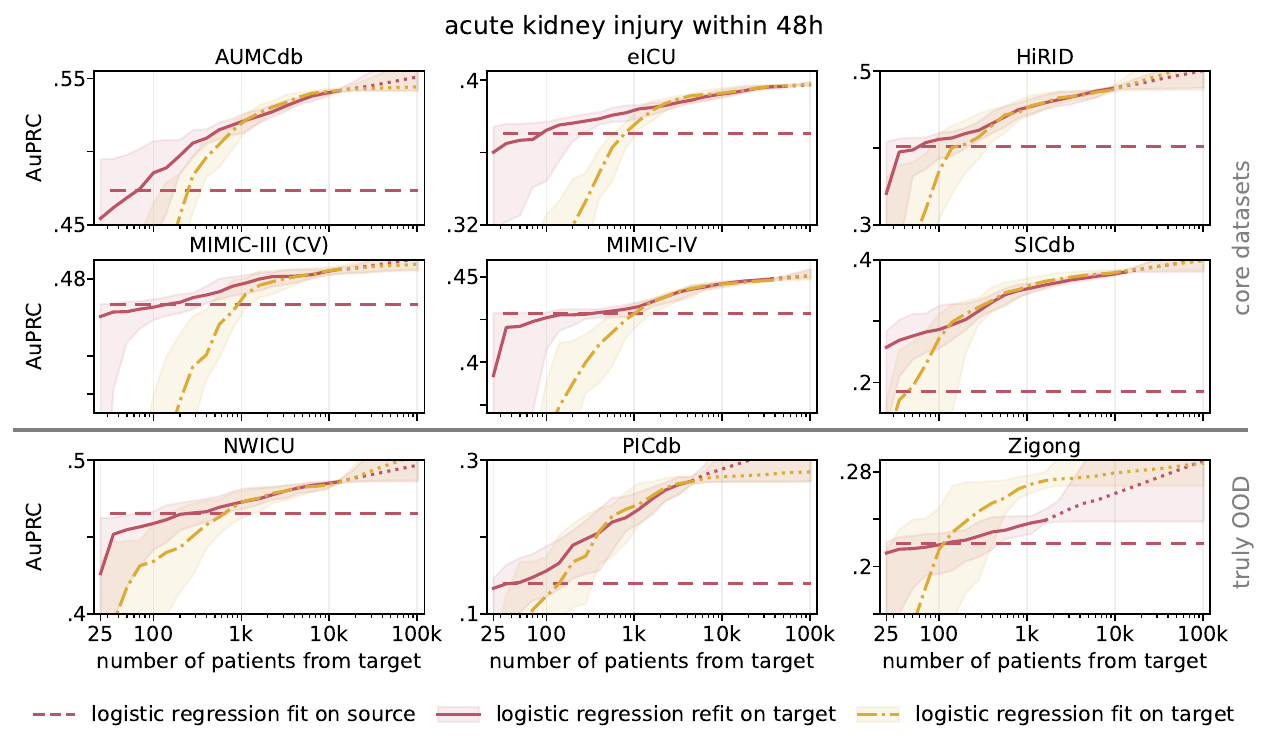}
    \caption{
        \label{fig:kidney_glm_n_samples}
        AuPRC (larger is better) predicting acute kidney injury within 48 hours as a function of available patients from the target dataset using logistic regression.
        Lines are medians and shaded areas are 80\% credible sets over 20 different subsampling seeds.
    }
\end{figure}

\begin{figure}[tbph]
    \centering
    \includegraphics[width=0.8\textwidth]{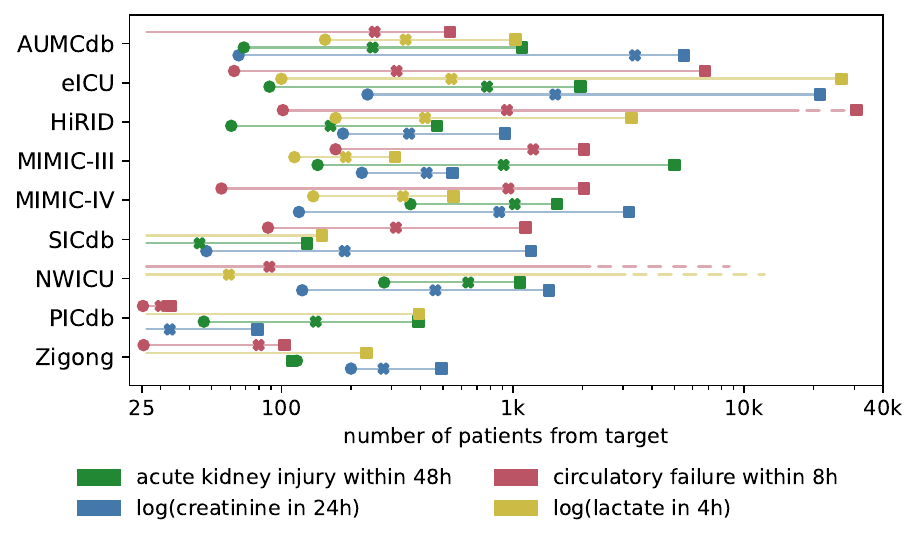}
    \caption{
        \label{fig:regimes_linear}
        Regime transitions for linear models as described in \cref{sec:methods:regimes} and \cref{fig:illustration}.
        \circlemarker~denotes the regime transition i $\to$ ii,
        \squaremarker~the regime transition ii $\to$ iii, and
        \xmarker~denotes the external data's value.
    }
\end{figure}

\FloatBarrier

\end{document}